\documentclass[12pt,letterpaper]{article}
\usepackage{jheppub}
\usepackage{epsfig}
\usepackage{amsmath}
\usepackage{amssymb}
\usepackage{subfigure}

\newcommand{\be}{\begin{equation}}
\newcommand{\ee}{\end{equation}}
\newcommand{\knl}{k_{\rm NL}}

\preprint{\hbox{CALT-68-2864}  }

\title{A Consistent Effective Theory of Long-Wavelength Cosmological Perturbations}
\author{Sean M. Carroll,}
\author{Stefan Leichenauer,}
\author{and Jason Pollack}
\affiliation{California Institute of Technology, Pasadena, CA 91125, U.S.A.}
\emailAdd{seancarroll@gmail.com, sleichen@theory.caltech.edu, jpollack@caltech.edu}

\abstract{Effective field theory provides a perturbative framework to study the evolution of cosmological large-scale structure. We investigate the underpinnings of this approach, and suggest new ways to compute correlation functions of cosmological observables. We find that, in contrast with quantum field theory, the appropriate effective theory of classical cosmological perturbations involves interactions that are nonlocal in time. We describe an alternative to the usual approach of smoothing the perturbations, based on a path-integral formulation of the renormalization group equations. This technique allows for improved handling of short-distance modes that are perturbatively generated by long-distance interactions.}

\begin{document}
\maketitle


\section{Introduction}
\label{sec-intro}

Effective Field Theory (EFT) has been successful in a wide variety of contexts. It allows a faithful description of physics at the length scales one is interested in measuring, without requiring detailed knowledge of dynamics at shorter distances. Instead the theory is formulated with an explicit length scale, the cutoff scale $\Lambda^{-1}$, below which modes can be nonlinear. The effects of short-wavelength physics appear only through parameters of the long-wavelength equations. The precise value of the cutoff scale is arbitrary and is chosen out of convenience, so physical quantities like correlation functions should not depend on it. Since $\Lambda$ enters calculations at intermediate stages, the requirement that the physics be $\Lambda$-independent non-trivially constrains the resulting long-wavelength theory.

Recently \cite{Baumann:2010tm,Carrasco:2012cv,Hertzberg:2012qn,Pajer:2013jj,Carrasco:2013sva,Mercolli:2013bsa,Carrasco:2013mua}, the principles of EFT have been applied to the problem of Large Scale Structure (LSS), the distribution of matter in the observable universe. Traditionally, standard perturbation theory (\cite{Fry:1983cj,Goroff:1986ep,Suto:1990wf,Makino:1991rp,Jain:1993jh,Scoccimarro:1996se}, reviewed in \cite{Bernardeau:2001qr}) has been used to compute the LSS correlation functions, but as has been emphasized in \cite{Baumann:2010tm,Carrasco:2012cv}, this is not a reliable basis for the theory. The problem is that for comoving wavenumbers $k> \knl \sim (10\ {\rm Mpc})^{-1}$ the perturbations have grown large enough that they become nonlinear. Standard perturbation theory (SPT), however, is only strictly valid when all modes remain linear, even those with $k>\knl$ where we know perturbation theory no longer applies. Starting at second order, SPT includes backreaction due to the propagation of modes of all wave numbers. So even for long-wavelength modes that have remained linear to a very good approximation, we should not trust the results of SPT at or beyond this order.

The alternative is to use EFT. The natural cutoff scale, $\Lambda \sim \knl$, is precisely the point where SPT breaks down. EFT treats modes with $k<\knl$ perturbatively, but does not attempt to make predictions about modes with $k>\knl$. Instead, within the EFT there are several additional parameters (beyond those found in SPT) whose values encode the effects that the nonlinear modes have on the linear modes. These parameters are not calculable within the EFT itself; they can be measured by experiment or extracted from N-body simulations, or (in principle) they may be calculated analytically from the full theory, which describes the behavior at all scales.\footnote{Of course, if we could perform such calculations, we wouldn't need to do perturbation theory at all.}

To formulate an EFT, one needs to have a model of the long-distance physics with parameters that can be adjusted to encode the effects of the unknown short-distance physics. For the problem of LSS, the full theory consists of a set of classical equations of motion which govern the evolution of the perturbations, together with a probability distribution over possible initial conditions at an early time $\tau_{\rm in}$. Given this data, we are tasked with computing the correlation functions of the long-wavelength perturbations at some arbitrary later time $\tau$, averaged over the ensemble of initial conditions. As we perform our analysis, we will work in as general a context as possible in order to clarify and extract the core concepts, so our perturbations will consist of an arbitrary number of fields and we will specify as little about the dynamics as possible.

In this paper we analyze carefully the derivation of the EFT of LSS. Along the way we uncover subtleties that distinguish this classical cosmological problem from the familiar context of quantum field theory. Unlike QFT, where loop diagrams arise from virtual particles, here they arise from integration over the probability distribution that specifies initial conditions. This distinction leads to important differences between the two problems.

We employ two separate methods of attack. First, we follow previous work and directly smooth the equations of motion in order to extract the effective long-distance evolution equations. Our strategy is to separate the long-wavelength and short-wavelength parts of the field, formally solve the short-wavelength equations of motion, and then plug the solution in to the long-wavelength equations of motion. This technique has been advocated in the previous works on the EFT of LSS cited above, but we carry out the procedure more completely and in greater generality. 

Our second technique is novel, based on a path-integral approach to the renormalization group. After expressing the sought-after correlation functions in terms of a partition function (a functional integral over initial configurations of the field), we use a modified version of Polchinski's renormalization group equations \cite{Polchinski:1983gv} to deduce the structure of the correlation functions of the long-wavelength modes.

We summarize our main results as follows:
\begin{itemize}
\item{In the smoothed-field approach, the effective equations of motion contain interactions which are nonlocal in time. We show that, at each order in perturbation theory, one can represent the effect of these nonlocal interactions in terms of local ones.}
\item{Smoothing the field does more than eliminate the nonlinear modes from the description: short-distance modes which are created perturbatively and remain small are also removed from the theory, leading to formal complications that will become numerically important at higher loop level if not properly accounted for.}
\item{The path integral approach, however, keeps the short-distance-yet-perturbative modes in the theory, allowing simpler formulas for the perturbative correlation functions. This makes this approach an attractive option for future development.}
\end{itemize}

The rest of the paper is organized as follows. First, we briefly explain in Section~\ref{sec-notation} the notation we will use for the remainder of the paper. In Section~\ref{sec-smoothing} we use the smoothing technique to extract the long-wavelength equations of motion from the full equations. We show how one can formally remove the short-distance field from the equations and construct a perturbative solution for the long-wavelength field alone. We identify several parameters in the long-wavelength effective theory which must be extracted from experiment. In Section~\ref{sec-pathintegral}, we use path integrals to solve the same problem in a new way. Polchinski's renormalization group is used to consistently determine the structure of the theory in terms of a set of integration kernels. In this formulation of the problem, these kernels represent the unknown parameters to be measured. We conclude and discuss future work in Section~\ref{sec-conclusion}. To provide a concrete example of our techniques, in Appendix~\ref{sec-pertcalc} we perform explicit calculations in the theory of LSS for an Einstein-de Sitter universe.

While this work was being completed, Ref.~\cite{Carrasco:2013mua} appeared which contains some overlapping discussion. In particular, its authors also discovered the need for interactions that are non-local in time.


\section{Notation}\label{sec-notation}

In this section we introduce the notation of the paper. Our main example is the theory of large-scale structure in a homogeneous FRW universe, for which the equations of motion are those of a pressureless fluid with a Newtonian gravitational interaction:
\begin{align}
0&= \partial_\tau \delta +\partial_j ((1+\delta)v^j)~,\\
0&=\partial_\tau v^i +\mathcal{H}v^i + \partial_i \Psi + v^j\partial_jv^i~.
\end{align}
Here the dynamical fields are the density perturbation $\delta \equiv \delta \rho/\rho$ and the velocity $v^i$. We work in conformal time $\tau$ with scale factor $a(\tau)$ and conformal Hubble parameter $\mathcal{H}=(\partial_\tau a)/a$. The potential $\Psi$ is related to $\delta$ through the Poisson equation, so it yields an interaction linear in $\delta$. These equations are approximate in the sense that they assume only a single matter component (namely pressureless dark matter), contain no relativistic corrections, and are valid on scales much smaller than the horizon. 

We immediately move to a set of equations general enough to account for all of these corrections:
\be\label{eq-general}
\mathcal{D}^i_j \phi^j - \frac{1}{2} M^i_{jk} \phi^j \phi^k - \frac{1}{3!} N^i_{jkl} \phi^i \phi^j \phi^l +\cdots = 0~.
\ee
We have collected all of our fields into a single object. In the particular case of standard perturbation theory, this is
\be
\phi^i(\tau) = \begin{pmatrix} \delta({\bf k}) \\ \theta({\bf k})\end{pmatrix},
\ee
where $\theta = \partial_i v^i$. (To the order at which we will work, vorticity can be ignored, so only the divergence of the velocity matters; our notation is sufficiently flexible that extension to more perturbation variables is immediate.) The Latin index labels both the species (in SPT, $\delta$ or $\theta$) and the wavenumber of the field, so (\ref{eq-general}) should be thought of as an equation in momentum space, obtained as the Fourier transform of the position space equation, and contraction of Latin indices denotes both a sum over species and an integral over wavenumber. 

We require that the linear operator $\mathcal{D}$ contains within it derivatives with respect to time of no higher than first order, but it can also contain non-derivative terms. Conservation of momentum (when it holds) is the statement that all interaction terms are proportional to $\delta$-functions:
\be
M^i_{jk} \propto \delta^{(3)}({\bf k}_i -{\bf k}_j - {\bf k}_k)
\ee
Other restrictions on the form of the coefficients will be imposed by rotational invariance or other symmetries of the problem. While this is an important aspect of the analysis, we will not focus on it here and so do not make special assumptions about the form of the interactions. Additionally, the interaction coefficients will in general be time-dependent, though we will often suppress this in the notation for simplicity. Note that the form of (\ref{eq-general}) guarantees that all interaction coefficients such as $M^i_{jk}$ and $N^i_{jkl}$ are symmetric in their lower indices.

In order to simplify the discussion, we will truncate the interactions at second order in the fields, so that
\be
\mathcal{D}^i_j \phi^j - \frac{1}{2} M^i_{jk} \phi^j \phi^k  = 0~.
\ee
This is not a fundamental limitation of our formalism, but a simplification made purely for clarity and notational convenience. It is trivial to extend our techniques and results to arbitrarily high-order interactions.\footnote{Note that we can incorporate higher-order time derivatives by increasing the number of fields (explicitly adding the velocity as an additional component of $\phi$, with an equation of motion setting it equal to the time derivative of the position, for example).}

To illustrate our techniques, we perform explicit calculations, which should clarify the meaning of the notation, in Appendix~\ref{sec-pertcalc}.


\section{The Smoothing Approach}\label{sec-smoothing}

In this section we carefully smooth the equations of motion to extract dynamics for the long-wavelength parts of the fields. The short-wavelength dynamics are formally solved (``integrated out") and plugged back into the long-wavelength equations. We formulate a perturbative expansion of the result, then use it to calculate correlation functions.

\subsection{Standard Perturbation Theory}\label{sub-spt}

The equations of motion that serve as our starting point are 
\be\label{eq-bare}
\mathcal{D}^i_j\phi^j - \frac{1}{2} M^i_{jk} \phi^j \phi^k = 0~.
\ee
Standard perturbation theory (SPT) calculates a perturbative expansion as a series in a formal parameter $\epsilon$,
\be\label{eq-phiexpand}
\phi_{\rm SPT}^i \equiv \epsilon   \phi_{(1)}^i +  \epsilon^2   \phi_{(2)}^i +  \epsilon^3  \phi_{(3)}^i + \cdots
\ee
then solves the equations of motion order by order in $\epsilon$. We are making a distinction here between the field $\phi$ and the formal series $\phi_{\rm SPT}$. The intention of SPT is to calculate $\phi$ by assuming that it is well approximated by $\phi_{\rm SPT}$, but, as discussed in the introduction, the approximation breaks down immediately for short-wavelength modes and at higher orders for long-wavelength modes (once backreaction is included). Ultimately we will argue that it is more appropriate to find a theory that is explicitly written purely in terms of the long-wavelength modes $\phi_{\rm L}$; the corresponding perturbation expansion appears in equation~(\ref{eq-phiLexpand}).

For completeness and future reference, we record here the equations of motion and solutions for SPT quantities up to $\mathcal{O}(\epsilon^3)$:
\begin{align}
\mathcal{O}(\epsilon):& &\mathcal{D}^i_j\phi_{(1)}^j &=0 ~, &  &\phi_{(1)}^i(\tau) = G^i_j(\tau ; \tau_{\rm in}) \phi_{\rm in}^j\,,
\label{phi1oftau}\\
\mathcal{O}(\epsilon^2):& &\mathcal{D}^i_j\phi_{(2)}^j &= \frac{1}{2} M^i_{jk} \phi_{(1)}^j \phi_{(1)}^k~, &  &\phi_{(2)}^i(\tau) = \frac{1}{2}\int_{\tau_{\rm in}}^\tau d\tau'~G^i_j(\tau ; \tau')  M^j_{kl}(\tau') \phi_{(1)}^k(\tau') \phi_{(1)}^l(\tau') \,,
\label{phi2oftau}\\
\mathcal{O}(\epsilon^3):& &\mathcal{D}^i_j\phi_{(3)}^j &= M^i_{jk} \phi_{(1)}^j \phi_{(2)}^k ~,&  &\phi_{(3)}^i(\tau) = \int_{\tau_{\rm in}}^\tau d\tau'~G^i_j(\tau ; \tau')  M^j_{kl}(\tau') \phi_{(1)}^k(\tau') \phi_{(2)}^l(\tau')\,.
\label{phi3oftau}
 \end{align}
\begin{figure*}[tb]
\begin{center}
\includegraphics[width=16 cm]{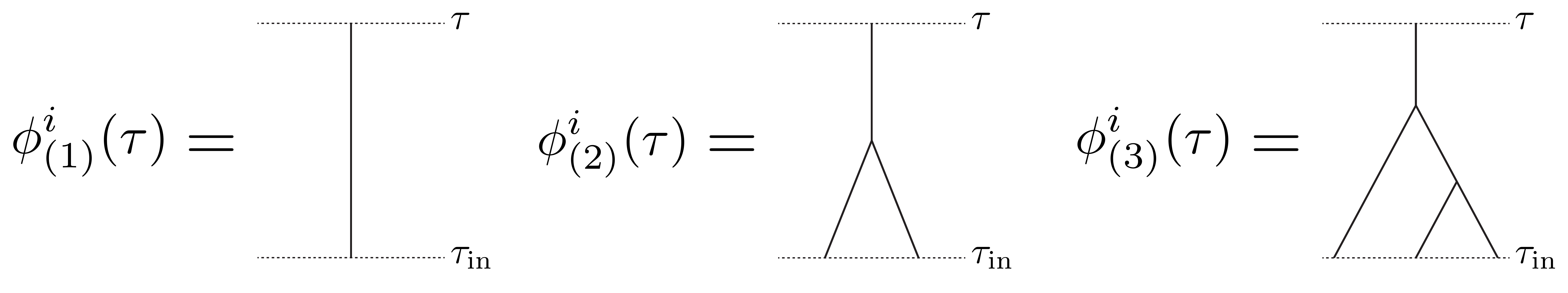}
\caption{Diagrammatic representation of the solution for $\phi^i_{(1)}(\tau)$,  $\phi^i_{(2)}(\tau)$, and $\phi^i_{(3)}(\tau)$, as given by equations (\ref{phi1oftau}), (\ref{phi2oftau}), and (\ref{phi3oftau}). Notation is as follows: vertical solid lines are associated with a Green function $G^i_j$. Vertices represent the interaction $M^i_{jk}$, and the position of each vertex is integrated over time. A solid line emerging from the bottom horizontal dotted line represents an initial condition $\phi^j_{\rm in}$, and the line reaching the upper horizontal dotted line is the quantity being calculated.}
\label{SPT-diagrams-fig}
\end{center}
\end{figure*}
Here $G^i_j$ represents the usual retarded Green function, which solves
\begin{align}
\mathcal{D}^i_j G^j_k(\tau;\tau_{\rm in})=\delta^i_k\delta(\tau-\tau_{\rm in}).
\end{align}
These solutions can be represented diagrammatically, as shown in Figure~\ref{SPT-diagrams-fig}.

Now we can compute correlation functions of the fields perturbatively by substituting the power series expansion (\ref{eq-phiexpand}). For the two-point function in particular, we have
\be\label{eq-SPTpowerspectrum}
\langle \phi_{\rm SPT}^i \phi_{\rm SPT}^j \rangle = \epsilon^2 \langle\phi_{(1)}^i\phi_{(1)}^j\rangle + \epsilon^4 \left[\langle\phi_{(1)}^i\phi_{(3)}^j\rangle+ \langle\phi_{(3)}^i\phi_{(1)}^j\rangle +\langle\phi_{(2)}^i\phi_{(2)}^j\rangle\right]+\cdots
\ee
Terms of order $\epsilon^n$ are expressed in terms of the $n$-point function of the initial conditions, which are specified at the initial time $\tau_{\rm in}$. For large scale structure, the initial time is usually taken as the time of matter-radiation equality so that matter domination can be assumed in the expression for the scale factor (which appears in the explicit form of the equations of motion). This is not important for us here, but it should be noted that $\tau_{\rm in}$ does not have to be ``the beginning" in any fundamental sense; it is merely the time at which we will begin calculating. We have have set all odd-point functions of the initial conditions to zero under the assumption that their distribution is Gaussian. This is only an approximation, since primordial non-Gaussianity as well as nonlinear effects prior to $\tau_{\rm in}$ will create non-Gaussianity at $\tau_{\rm in}$. However, it should be a numerically good approximation to ignore such effects, and in any case corrections of this type are easily incorporated into the formalism.

For illustration, we will compute one of the terms in (\ref{eq-SPTpowerspectrum}):
\be\begin{split}
\langle\phi_{(1)}^i\phi_{(3)}^j\rangle &= \langle \phi^i_{(1)}(\tau) \int_{\tau_{\rm in}}^\tau d\tau'~G^j_k(\tau ; \tau')  M^k_{lm}(\tau') \phi_{(1)}^l(\tau') \phi_{(2)}^m(\tau')\rangle\\
&=\frac{1}{2} \int_{\tau_{\rm in}}^\tau d\tau'\int_{\tau_{\rm in}}^{\tau'} d\tau''~~G^j_k(\tau ; \tau')  M^k_{lm}(\tau')G^m_n(\tau' ; \tau'')  M^n_{op}(\tau'')  \\
& \qquad \qquad\qquad\qquad \times \langle \phi^i_{(1)}(\tau) \phi_{(1)}^l(\tau')\phi_{(1)}^o(\tau'') \phi_{(1)}^p(\tau'') \rangle~.
\end{split}
\ee
Assuming Gaussianity, Wick's theorem can be used to evaluate the four-point function appearing here. Momentum conservation together with the explicit form of the interaction coefficients $M$ leads to some simplifications. As discussed in Appendix~\ref{sec-pertcalc}, there are two non-vanishing Wick contractions which contribute to this correlation function, and they give equal contributions to the total. In slightly expanded notation, the result is given in (\ref{eq-P13}), reproduced here:
\begin{figure*}[tb]
\begin{center}
\includegraphics[width=12 cm]{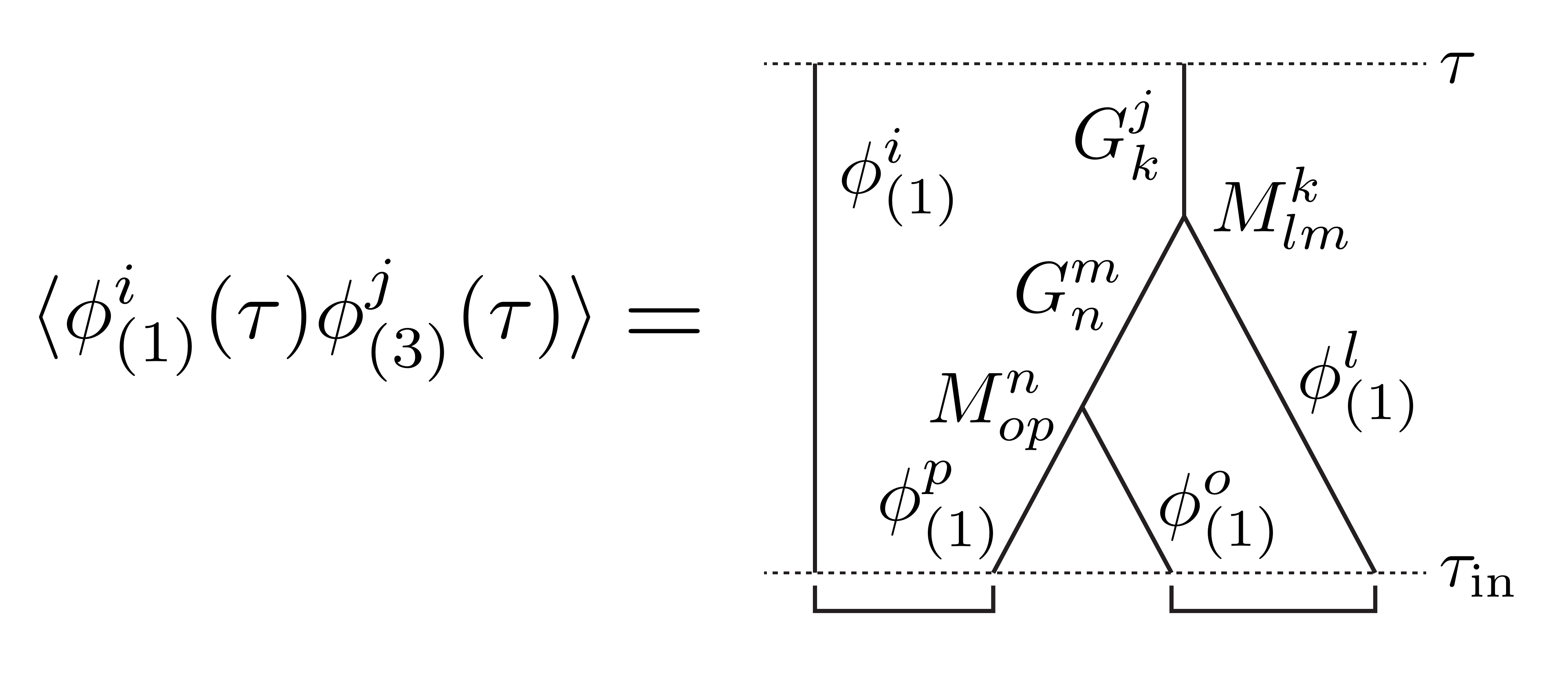}
\caption{Diagrammatic representation of the correlator $\langle \phi^i_{(1)} \phi^j_{(3)}\rangle$,
as expressed in equation (\ref{eq-P13copy}). Here we have explicitly indicated the quantities associated with each line and vertex. The bottom brackets represent contraction of the two lines, which is carried out by summing with the linear power spectrum. Momentum in the loop labeled with indices $l, m, n, o$ is integrated over.
The other possible contraction, linking $\phi^o_{(1)}$ and $\phi^p_{(1)}$, vanishes in the theory of LSS.}
\label{P13-fig}
\end{center}
\end{figure*}
\be\label{eq-P13copy}
\begin{split}
\langle \phi^i_{(1)} \phi^j_{(3)}\rangle& =(2\pi)^3 \delta^{(3)}({\bf k}_i + {\bf k}_j)\\
&\qquad \times \int_{\tau_{\rm in}}^\tau d\tau' \int_{\tau_{\rm in}}^{\tau'} d\tau'' \int\frac{d^3q}{(2\pi)^3} G^j_k(\tau; \tau')M^k_{lm}({\bf k}_j;{\bf q},{\bf k}_j-{\bf q}) \\
& \qquad\qquad \times G^m_n(\tau';\tau'') M^n_{op}({\bf k}_j-{\bf q};-{\bf q},{\bf k}_j)
P_{(11)}^{ip}(\tau,\tau''|k_i)P_{(11)}^{lo}(\tau',\tau''|q) ~.
\end{split}
\ee
This is shown diagrammatically in Figure~\ref{P13-fig}.
Here we have made use of the linear power spectrum $P_{(11)}^{ij}$, defined through the equation
\be
\langle \phi^i_{(1)}(\tau_1) \phi^j_{(1)}(\tau_2) \rangle = G^i_l(\tau_1;\tau_{\rm in}) G^j_m(\tau_2; \tau_{\rm in})\langle \phi^l_{\rm in} \phi^m_{\rm in} \rangle\equiv P_{(11)}^{ij}(\tau_1,\tau_2 | k_i)(2\pi)^3 \delta^{(3)}({\bf k}_i + {\bf k}_j)~.
\ee
This is a one-loop expression, where the name comes from the loop which appears in its diagrammatic representation. After momentum conservation is imposed at each vertex, a single integration over momentum remains (${\bf q}$ in our formula). That is the loop momentum.

At the end of the calculation, $\epsilon$ is set equal to one to obtain the actual solution, and the justification for the expansion is that the field itself is small. Then the nonlinear terms in the equation of motion are small compared to the linear terms and the higher order corrections $\phi_{(n)}$ for $n\geq 2$ systematically take them into account. However, even if the initial conditions are small, it may be that the dynamics causes the field value to grow with time. In the theory of large scale structure, the perturbations with $k > \knl$ have become large by the present, and so (\ref{eq-P13copy}) is no longer a small perturbation. This is the point at which SPT breaks down. 


\subsection{Effective Equations of Motion}

We turn next to an effective theory which can incorporate the nonlinear interactions while still maintaining perturbativity. The fields are still expanded in a power series, but one that is conceptually different from (\ref{eq-phiexpand}) of SPT.

We begin by splitting $\phi$ at the cutoff scale $\Lambda$, dividing it into a long-wavelength piece, $\phi_{\rm L}$, and a short-wavelength piece, $\phi_{\rm S}$, so that $\phi =\phi_{\rm L} + \phi_{\rm S}$. The split is accomplished using a smoothing function, $W_\Lambda$, which extracts $\phi_{\rm L}$ from the fundamental field $\phi$. In position space, we would smooth the field via convolution:
\be
\phi_{\rm L}(x) = \int d^3y~ W_\Lambda(x-y) \phi(y)~.
\ee
Under Fourier transform, convolution becomes multiplication. We will denote the Fourier transform of $W_\Lambda$ also by $W_\Lambda$, but there should be no confusion since we work almost entirely in momentum space. Then we have
\be
 \phi_{\rm L}^i = W_\Lambda(i)\phi^i~.
\ee
We have written $W_\Lambda$ as a function of the index $i$ of the field because it is a function of momentum, which is part of that index. There is no implied sum on $i$ in this formula. The properties of $W_\Lambda$ are somewhat arbitrary, but we will find it most convenient to use $W_\Lambda({\bf k}) = \Theta(\Lambda - |{\bf k}|)$. In Section~\ref{sec-pathintegral} we will find it convenient for $W_\Lambda$ to be differentiable, so a smoothed version of the $\Theta$-function is more appropriate.\footnote{It should be noted that, unless $W_\Lambda$ and $1-W_\Lambda$ have disjoint support, there is not a clear distinction between long-wavelength and short-wavelength fields. This leads to complications that we will ignore in this section. As long as $W_\Lambda$ differs from a $\Theta$-function only in a small neighborhood around $\Lambda$ (to allow for a smooth transition between $0$ and $1$, if desired), the numerical error from this approximation will be arbitrarily small. In \cite{Baumann:2010tm} and its successors, a Gaussian form for $W_\Lambda$ was assumed. This choice of smoothing is invertible, and therefore in principle retains information about short-distance physics, contrary to the spirit of EFT.} 

For simplicity, we will assume that the linear part of the equation of motion, $\mathcal{D}^i_j$, is diagonal in momentum space (as is the case for LSS), so that the smoothed equation of motion is
\begin{align}
0&= \mathcal{D}^i_j\phi_{\rm L}^j - \frac{1}{2} W_\Lambda(i)M^i_{jk} \phi^j \phi^k~\\
&= \mathcal{D}^i_j\phi_{\rm L}^j - \frac{1}{2} W_\Lambda(i)M^i_{jk} \phi_{\rm L}^j \phi_{\rm L}^k - W_\Lambda(i)M^i_{jk}\phi_{\rm S}^j \phi_{\rm L}^k - \frac{1}{2} W_\Lambda(i)M^i_{jk} \phi_{\rm S}^j \phi_{\rm S}^k~.\label{eq-long}
\end{align}
The first two terms represent interactions among long-wavelength fields producing the long-wavelength field, while the remaining terms are interactions of the long-wavelength field with the short-wavelength field. Subtracting (\ref{eq-long}) from (\ref{eq-bare}) we find
\be\label{eq-short}
\mathcal{D}^i_j\phi_{\rm S}^j - \frac{1}{2} (1-W_\Lambda(i))M^i_{jk} \phi_{\rm S}^j \phi_{\rm S}^k - (1-W_\Lambda(i))M^i_{jk}\phi_{\rm L}^j \phi_{\rm S}^k - \frac{1}{2} (1-W_\Lambda(i))M^i_{jk} \phi_{\rm L}^j \phi_{\rm L}^k~.
\ee
Treating $\phi_{\rm L}$ formally as a background field, specified for all times, we can solve (\ref{eq-short}) for the short-wavelength field as a functional of the long-wavelength field, $\phi_{\rm S}[\phi_{\rm L}]$. This result can then be substituted in for $\phi_{\rm S}$ in (\ref{eq-long}) to obtain an equation for $\phi_{\rm L}$ alone.

\begin{figure*}[tb]
\begin{center}
\includegraphics[width=14 cm]{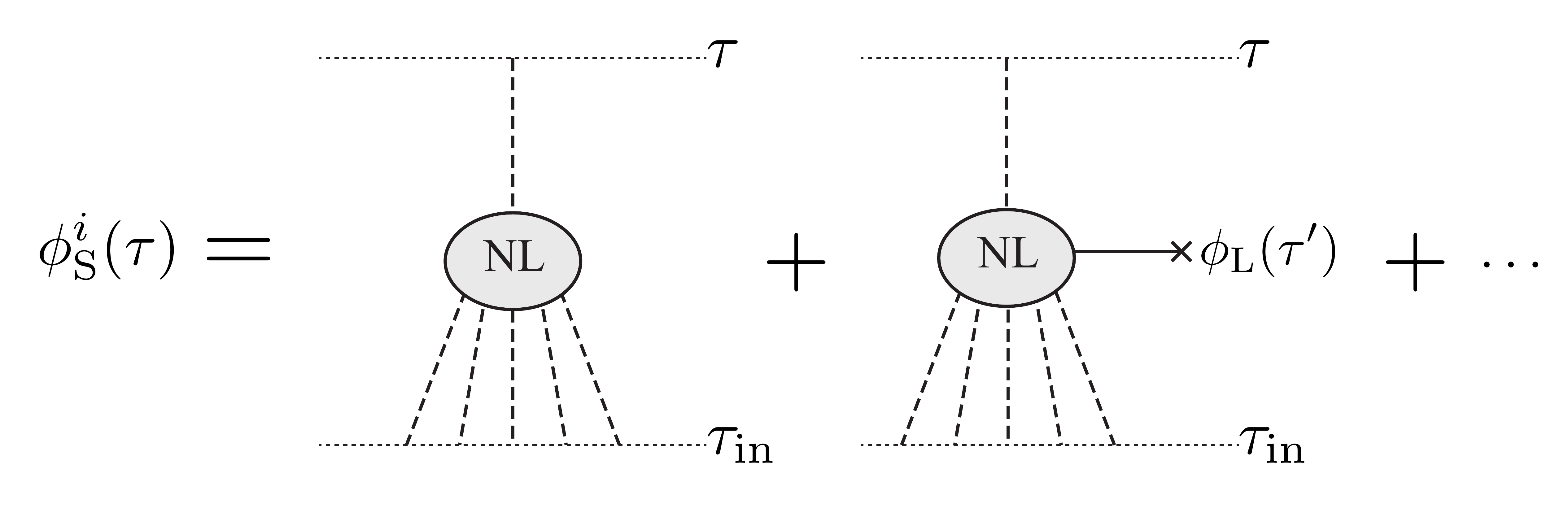}
\caption{Diagrammatic representation of the Taylor expansion for $\phi^i_{\mathrm{S}}$ considered as a
functional of $\phi^j_{\mathrm{L}}$, as expressed in equation (\ref{eq-shortpert}). Dashed lines represent (arbitrary numbers of) the field $\phi_{\rm S}$, and the NL blob stands for nonlinear interactions. In the second diagram we see the effects of the background field $\phi_{\rm L}$, thought of as an external source.}
\label{phiS-fig}
\end{center}
\end{figure*}

More concretely, we expand the functional $\phi_{\rm S}[\phi_{\rm L}]$ in a Taylor series about its value when the long-wavelength modes are set equal to zero, $\phi_{{\rm S}0}\equiv{\phi_{\rm S}[\phi_{\rm L}=0]}$:
\be\label{eq-shortpert}
\phi_{\rm S}^i(\tau) = \phi_{{\rm S}0}^i(\tau) + \int_{\tau_{\rm in}}^\tau d\tau' ~ \left.\frac{\partial \phi_{\rm S}^i(\tau)}{\partial \phi_{\rm L}^j(\tau')}\right|_{\phi_{\rm L}=0}\phi_{\rm L}^j(\tau') + \cdots
\ee
This formula is expressed diagrammatically in Figure~\ref{phiS-fig}.
The first term is the solution of (\ref{eq-short}) obtained by setting $\phi_{\rm L}=0$ for all times. The second term is the first correction coming from the incorporation of non-zero $\phi_{\rm L}$, where we think of $\phi_{\rm L}$ as an arbitrary background specified at all times. We stress that the functional derivative $\partial \phi_{\rm S}/\partial \phi_{\rm L}$ is being evaluated on the configuration $\phi_{\rm L}=0$; we have explicitly indicated this in the equation above, but for brevity will suppress it in the rest of the paper. The limits of integration are the time when initial conditions are specified, $\tau_{\rm in}$, and the time at which $\phi_{\rm S}$ is being evaluated, $\tau$; causality requires that field values at times beyond $\tau$ do not contribute.

Up to this point, we have assumed that the initial conditions at $\tau_{\rm in}$ are completely specified. For the application to LSS, however, we have a probability distribution over initial conditions.  For now we imagine selecting one particular initial condition from the ensemble; the average over all possibilities is only taken at the very end of the calculation. If instead we take the expectation value over the short-wavelength initial conditions at this intermediate stage, we will miss some correlations. This effect may become non-negligible at high orders in perturbation theory.

Returning to (\ref{eq-long}), we can plug in our perturbative expansion (\ref{eq-shortpert}) to get
\be\label{eq-long2}
\begin{split}
0=& ~\mathcal{D}^i_j\phi_{\rm L}^j - \frac{1}{2} W_\Lambda(i)M^i_{jk} \phi_{\rm L}^j \phi_{\rm L}^k - W_\Lambda(i)M^i_{jk}\phi_{{\rm S}0}^j \phi_{\rm L}^k - \frac{1}{2} W_\Lambda(i)M^i_{jk} \phi_{{\rm S}0}^j \phi_{{\rm S}0}^k\\
& -  W_\Lambda(i)M^i_{jk} \phi_{{\rm S}0}^j \int^\tau\frac{\partial \phi_{\rm S}^k(\tau)}{\partial \phi_{\rm L}^l(\tau')}\phi_{\rm L}^l(\tau')- W_\Lambda(i)M^i_{jk}\phi_{\rm L}^j\int^\tau\frac{\partial \phi_{\rm S}^k(\tau)}{\partial \phi_{\rm L}^l(\tau')}\phi_{\rm L}^l(\tau')+\cdots~,
\end{split}
\ee
where the $\cdots$ represent terms that, as we will see below, are higher order.

Notice that the induced interactions of the long-wavelength field with itself are non-local in time. This is a very important conceptual point. The modes we have integrated out are short-distance modes, but depending on the dynamics they may be long-lived. This means that the functional derivative $\partial \phi_{\rm S}^k(\tau)/\partial \phi_{\rm L}^l(\tau')$ may have significant support even when $\tau$ and $\tau'$ are very different. There are different possible strategies for dealing with these terms. We will see below how they can be systematically accounted for in perturbation theory. These interaction terms, and their perturbative forms, represent the new parameters needed to define the EFT. 


\subsection{Effective Perturbation Theory}

The next step, as in section \ref{sub-spt} above, is to formally expand $\phi_{\rm L}$ in a parameter $\epsilon$ and use perturbation theory on this new, effective equation. We will assume that some perturbative description is valid where $\phi_{\rm L} \approx \phi_{(1)}$ is still true to leading order (for the long-wavelength modes), and so the new terms should only give corrections to that. In order to make progress, we need to decide how many powers of $\epsilon$ to assign to the new terms generated by interactions involving $\phi_{\rm S}$. This turns out to be an involved question, and we will need to make use of some assumptions about the dynamics. For concreteness, we will specifically refer to the theory of LSS. 
  
The short-wavelength field $\phi_S$ is a complicated object. First consider $\phi_{{\rm S}0}$, the solution for the short-wavelength perturbation in the absence of long-wavelength perturbations. For modes near the cutoff, the linear perturbation theory should still be approximately valid and $\phi_{{\rm S}0} \approx \phi_{(1)}$ should hold, so that $\phi_{{\rm S}0} \sim \epsilon$.  For the truly nonlinear modes, we can no longer assume that $\phi_{{\rm S}0}$ is small. However, because perturbation theory is still valid at long wavelengths, we will assume that the order of magnitude of the effects of the modes at more nonlinear scales is well-estimated by the effects of the modes at only slightly nonlinear scales. In other words, the scaling of the interaction terms we get by assuming $\phi_{{\rm S}0} \sim \epsilon$ will be assumed to be the correct scaling. This is an important assumption and we have not proved it. More detailed analyses of the order-of-magnitude of nonlinear effects within the theory of LSS are performed in \cite{Carrasco:2012cv,Hertzberg:2012qn}. There it is shown that the nonlinear effects are under control and that this estimation is ultimately correct. However, this point may be an important restriction on the general applicability of EFT methods to arbitrary equations of motion.

Similar comments can be made about the functional derivative $\partial \phi_{\rm S} / \partial \phi_{\rm L}$. However, at the level of the second derivative, $\partial^2 \phi_{\rm S}/\partial \phi_{\rm L}^2$, there is a new effect. Because two long-wavelength modes can come together to make a short-wavelength mode through the interaction $M$, the second derivative $\partial^2 \phi_{\rm S}/\partial \phi_{\rm L}^2$ will have a term that does not contain any factors of $\epsilon$. A term in $\partial^2 \phi_{\rm S}/\partial \phi_{\rm L}^2$ with zero powers of $\epsilon$ thus acts at the same order in perturbation theory as a term in $\partial \phi_{\rm S}/\partial \phi_{L}$ with a single power of $\epsilon$. This complicates the perturbation expansion.

In the specific case of LSS, the origin of this complication is that the smoothing procedure has removed too many modes from the theory.  A short-wavelength mode that is created from two long-wavelength modes, produced by the unsuppressed term in $\partial^2 \phi_{\rm S}/\partial \phi_{\rm L}^2$, is not a nonlinear mode. The reason is that short-wavelength modes created in this way are nearly identical to long-wavelength modes created in the same way: they are initially small (i.e., linear), and grow according to a ${\rm \bf k}$-independent linear growth function. Like these long-wavelength modes, we expect short-wavelength modes generated in this way to remain linear.

To summarize, in the expansion of $\phi_{\rm S}$ there will be two types of terms. First, there will be terms representing short-wavelength modes which have dynamically evolved from short-wavelength initial conditions. We will assume that all such modes give $\mathcal{O}(\epsilon)$ contributions to $\phi_{\rm S}$ and each of its functional derivatives. Second, there will be terms representing short-wavelength modes which are generated through the action of the background field $\phi_{\rm L}$ alone. These terms do not represent nonlinear physics, and it is a defect of the formalism that they appear here as modes to be smoothed over. We will ignore these terms completely for now, even though they contribute non-$\epsilon$-suppressed contributions to the functional derivative. One reason why it is possible to ignore them is because they make up only a small portion of the phase space at low order: when only a few long-wavelength modes combine together to make a short-wavelength mode, the resulting wavenumber will not be much greater than the cutoff $\Lambda$. At higher orders, however, when there are more long-wavelength fields propagating, the numerical error caused by a failure to systematically account for these modes will be greater. We leave incorporation of these higher-order effects into this formalism for future work, although we will have a bit more to say on the topic below when we discuss correlation functions. In Section \ref{sec-pathintegral} below we present an alternative formalism that naturally takes these modes into account.

We will now expand $\phi_{\rm L}$ as a power series in $\epsilon$, as we did with $\phi$ in (\ref{eq-phiexpand}). To make the dependence on nonlinear effects explicit, we decompose each order in the expansion into a ``standard'' piece and a piece generated by short-distance physics. We write
\be
\phi_{\rm L}^i = \epsilon (\phi_{{\rm L}(1)}^i + \Delta \phi_{{\rm L}(1)}^i) +  \epsilon^2  (\phi_{{\rm L}(2)}^i+\Delta \phi_{{\rm L}(2)}^i) +  \epsilon^3 (\phi_{{\rm L}(3)}^i +\Delta \phi_{{\rm L}(3)}^i) + \cdots
\label{eq-phiLexpand}
\ee
The functions $\phi_{{\rm L}(n)}$ are defined to be the same as the $\phi_{(n)}$ as in standard perturbation theory, with the replacement $M^i_{jk} \to W_\Lambda(i)M^i_{jk}$, and using the smoothed long-wavelength initial conditions $W_\Lambda(i)\phi_{\rm in}$ instead of the unsmoothed initial conditions. Equivalently, one could make the replacement $G^i_j \to W_\Lambda(i) G^i_j$, making use of the assumption that $G$ is diagonal in momentum. Physically, this means that only long-wavelength fields are allowed to propagate in the construction of the $\phi_{{\rm L}(n)}$. Diagrammatically, all lines have their momenta cut off. 

The effects of the short-wavelength interactions are denoted by $\Delta \phi_{{\rm L}(n)}$. We can find equations of motion for the $\Delta \phi_{\rm L}$ terms by expanding the effective equation of motion in $\epsilon$. At $\mathcal{O}(\epsilon)$ we find
\be
\mathcal{D}^i_j \Delta \phi_{{\rm L}(1)}^j =0 ~.
\ee
Since the initial conditions are already accounted for in $\phi_{{\rm L}(1)}$, the solution to this equation is $\Delta\phi_{{\rm L}(1)}=0$. This is just a reflection of our assumption that the effects of the short-distance scales on the long-distance scales can be treated perturbatively. 

\begin{figure*}[tb]
\begin{center}
\includegraphics[width=16 cm]{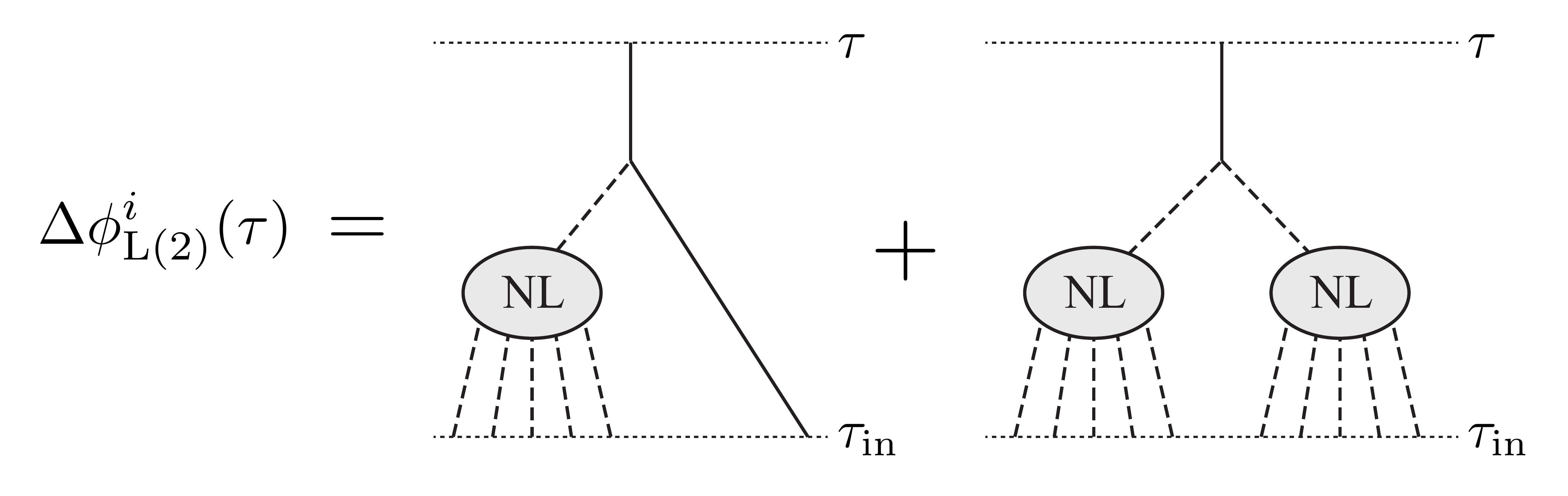}
\caption{Diagrammatic representation of the solution for $\Delta\phi^i_{\mathrm{L}(2)}$,
as expressed in equation (\ref{eq-delta-phi-L-2}). Dashed lines are the short-wavelength field $\phi_{{\rm S}0}$, while solid lines are the long-wavelength field $\phi_{\rm L}$. As in Figure~\ref{phiS-fig}, NL blobs represent nonlinear interactions.}
\label{DeltaphiL2-fig}
\end{center}
\end{figure*}

Using this result, we find at $\mathcal{O}(\epsilon^2)$
\be
\mathcal{D}^i_j\Delta\phi_{{\rm L}(2)}^j = W_\Lambda(i)M^i_{jk} \phi_{{\rm S}0}^j \phi_{{\rm L}(1)}^k+\frac{1}{2} W_\Lambda(i)M^i_{jk} \phi_{{\rm S}0}^j \phi_{{\rm S}0}^k~.
\ee
Here we see the first effects of the short-distance physics. We can easily write down the solution for $\Delta\phi^i_{{\rm L}(2)}$ from this equation:
\be
\Delta \phi^i_{{\rm L}(2)}(\tau) = \int_{\tau_{\rm in}}^\tau d\tau' ~G^i_j(\tau;\tau')\left[ W_\Lambda(j)M^j_{kl} \phi_{{\rm S}0}^k \phi_{{\rm L}(1)}^l+\frac{1}{2} W_\Lambda(j)M^j_{kl} \phi_{{\rm S}0}^k \phi_{{\rm S}0}^l\right]~.
\label{eq-delta-phi-L-2}
\ee
This solution is expressed diagrammatically in Figure~\ref{DeltaphiL2-fig}.

At $\mathcal{O}(\epsilon^3)$ we have 
\be
\mathcal{D}^i_j \Delta\phi_{{\rm L}(3)}^j = W_\Lambda(i)M^i_{jk} \phi_{{\rm S}0}^j (\phi_{{\rm L}(2)}^k+\Delta\phi_{{\rm L}(2)}^k)+ W_\Lambda(i)M^i_{jk} (\phi_{{\rm S}0}^j + \phi_{{\rm L}(1)}^j) \int^\tau\frac{\partial \phi_{\rm S}^k(\tau)}{\partial \phi_{\rm L}^l(\tau')}\phi_{{\rm L}(1)}^l(\tau') ~.
\ee
Here at third order we find the first nonlocal-in-time interactions. However, in the context of this perturbative expansion it is easy to deal with them. Since we already know the full time-dependence of $\phi_{{\rm L}(1)}$, we can factor it out of the time integral:
\be
\int^\tau_{\tau_{\rm in}}d\tau'~\frac{\partial \phi_{\rm S}^k(\tau)}{\partial \phi_{\rm L}^l(\tau')}\phi_{{\rm L}(1)}^l(\tau')= \left[ \int^\tau_{\tau_{\rm in}}d\tau'~\frac{\partial \phi_{\rm S}^k(\tau)}{\partial \phi_{\rm L}^l(\tau')}\left[G(\tau; \tau')^{-1}\right]^l_m\right] \phi_{{\rm L}(1)}^m(\tau)~.
\ee
(The propagator is guaranteed to be invertible, because the equations of motion are reversible.) Because perturbation theory is a recursive process, this procedure can be repeated at each order. Then perturbatively the nonlocal-in-time interaction is expressed in terms of local-in-time interactions. We can now write a simple expression for $\Delta \phi_{{\rm L}(3)}$:
\be\begin{split}
\Delta \phi^i_{{\rm L}(3)}(\tau) =& \int_{\tau_{\rm in}}^\tau d\tau' ~G^i_j(\tau;\tau')  \bigg[ W_\Lambda(j)M^j_{kl} \phi_{{\rm S}0}^k \left(\phi_{{\rm L}(2)}^l+\Delta\phi_{{\rm L}(2)}^l\right) \\
& +W_\Lambda(j)M^j_{kl} \left(\phi_{{\rm S}0}^k + \phi_{{\rm L}(1)}^k\right) \left[ \int^{\tau'}_{\tau_{\rm in}}d\tau''~\frac{\partial \phi_{\rm S}^l(\tau')}{\partial \phi_{\rm L}^m(\tau'')}\left[G(\tau'; \tau'')^{-1}\right]^m_n\right] \phi_{{\rm L}(1)}^n(\tau')\bigg]~.
\end{split}
\label{eq-deltaphi3}
\ee


\subsection{Correlation Functions}

Using the formalism above we can compute correlation functions of the long-wavelength field $\phi_{\rm L}$. The power spectrum in particular can be written as
\be\begin{split}
\langle \phi_{\rm L}^i \phi_{\rm L}^j \rangle 
=& \langle\phi_{{\rm L}(1)}^i\phi_{{\rm L}(1)}^j\rangle + \langle\phi_{{\rm L}(1)}^i\phi_{{\rm L}(3)}^j\rangle+ \langle\phi_{{\rm L}(3)}^i\phi_{{\rm L}(1)}^j\rangle +\langle\phi_{{\rm L}(2)}^i\phi_{{\rm L}(2)}^j\rangle +  \langle\phi_{{\rm L}(1)}^i \Delta \phi_{{\rm L}(3)}^j\rangle \\
&+ \langle\Delta \phi_{{\rm L}(3)}^i\phi_{(1)}^j\rangle+\langle\phi_{{\rm L}(2)}^i \Delta \phi_{{\rm L}(2)}^j\rangle+ \langle \Delta\phi_{{\rm L}(2)}^i \phi_{{\rm L}(2)}^j\rangle+\langle \Delta\phi_{{\rm L}(2)}^i\Delta\phi_{{\rm L}(2)}^j\rangle + \cdots~.
\end{split}
\label{eq-phiLpowerspectrum}
\ee
The first line is very similar to the 1-loop result from SPT, except every propagating field is long-wavelength. The second line represents corrections to that result up to the same order. Note that the sum of both lines is meant to be $\Lambda$-independent. In practice, the $\Lambda$-dependence of the $\Delta\phi$ terms is determined by this requirement since they cannot be calculated from first principles. An example of such a calculation is performed in Appendix~\ref{sec-pertcalc}.

We note one technical point about this calculation. When computing the SPT-like terms in the EFT expansion of the correlation function, every single field is supposed to be restricted to being long-wavelength. In general, this will impose multiple different cutoffs on each loop integral beyond those imposed by removing the short-wavelength initial conditions, since some of the lines within the loop carry sums of other loop momenta and the external momentum. This technical annoyance has, as far as we are aware, gone unnoticed in the literature. The effects of such a restriction will be small, subleading in $k_{\rm external}/\Lambda$ for low orders in perturbation theory, but more important at higher orders. This is related to the issue we mentioned above regarding the contributions to $\phi_{\rm S}$ which are not actually nonlinear. Those contributions precisely correspond to lines in the loop diagram which go over the cutoff even when all initial conditions are long-wavelength only. It would be preferable to have a formalism where this was accounted for automatically, and the only cutoff that had to be performed on the diagram was a cutoff on the initial conditions. This is achieved by the path integral formalism discussed in the next section.


\section{A Path Integral Approach}\label{sec-pathintegral}

In this section we use a statistical path integral to incorporate the effects of the probability distribution over initial conditions on the equations of motion. By following the Polchinski renormalization group procedure \cite{Polchinski:1983gv} we can deduce the structure of the effective theory. Instead of smoothing the equations of motion, we will demand that the coefficients in the effective action are closed under renormalization group flow. We find general results that agree with the analysis of the previous section, up to the issues relating to linear portions of $\phi_S$ discussed above, and offer new insight into the effective theory.


\subsection{Polchinski RG}\label{sec-RGreview}

In this section we review the usual equations of the Polchinski RG using our condensed notation, where a Latin index is both a discrete label for field species and also a continuous label for momentum. Consider an action of the form
\be
S(\phi, \Lambda)  = -\frac{1}{2} \phi^i [P(\Lambda)^{-1}]_{ij} \phi^j + S_{\rm int}(\phi,\Lambda)~,
\ee
where $P(\Lambda)$ is a symmetric matrix which depends on the momentum cutoff $\Lambda$.\footnote{$P$ should be thought of as a smoothly-varying function of momentum $k$ and the cutoff $\Lambda$ which is $\Lambda$-independent when $k <\Lambda$ and zero when $k > \Lambda$, and transitions rapidly between these two phases for $k\sim \Lambda$. In the EFT of LSS, $P$ is the smoothed initial power spectrum.}
The partition function,
\be
Z = \int \mathcal{D}\phi~ e^{S}~,
\ee
is a physical object and so should be independent of $\Lambda$, which was introduced artificially. Taking the $\Lambda$-derivative gives
\be
\frac{dZ}{d\Lambda} = \int \mathcal{D}\phi \left(-\frac{1}{2} \phi^j \frac{d}{d\Lambda}\left[P(\Lambda)^{-1}\right]_{ij} \phi^j + \frac{d}{d\Lambda}S_{\rm int}(\phi,\Lambda)\right) e^S~.
\ee
Then consider the quantity
\be
\begin{split}
\frac{\partial}{\partial \phi^i}&\left( \frac{dP^{ij}}{d\Lambda}[P^{-1}]_{jk}\phi^ke^{S} + \frac{1}{2}\frac{dP}{d\Lambda}^{ij}\frac{\partial S}{\partial \phi^j}e^{S} \right) \\
&=\left(\frac{dP^{ij}}{d\Lambda}[P^{-1}]_{ji}+ \frac{dP^{ij}}{d\Lambda}[P^{-1}]_{jk}\phi^k  \frac{\partial S}{\partial \phi^i} + \frac{1}{2}\frac{dP}{d\Lambda}^{ij}\frac{\partial S}{\partial \phi^j}\frac{\partial S}{\partial \phi^i} +\frac{1}{2} \frac{dP}{d\Lambda}^{ij}\frac{\partial^2 S}{\partial \phi^i \partial \phi^j}  \right)e^{S} \\
&=\left(\frac{1}{2}\frac{dP^{ij}}{d\Lambda}[P^{-1}]_{ji}+ \frac{1}{2}\phi^i\frac{d}{d\Lambda}[P^{-1}]_{ij}\phi^j+ \frac{1}{2}\frac{dP}{d\Lambda}^{ij}\frac{\partial S_{\rm int}}{\partial \phi^j}\frac{\partial S_{\rm int}}{\partial \phi^i} + \frac{1}{2}\frac{dP}{d\Lambda}^{ij}\frac{\partial^2 S_{\rm int}}{\partial \phi^i \partial \phi^j}  \right)e^{S} ~.
\end{split}
\ee
Since this is a total derivative, it will vanish when (functionally) integrated with respect to $\phi$. Then, up to a field-independent but $\Lambda$-dependent shift in the action, the partition function will be independent of $\Lambda$ if
\be\label{eq-actionrg}
\frac{d}{d\Lambda}S_{\rm int}(\phi,\Lambda) = -\frac{1}{2}\left(\frac{dP}{d\Lambda}^{ij}\frac{\partial S_{\rm int}}{\partial \phi^j}\frac{\partial S_{\rm int}}{\partial \phi^i} + \frac{dP}{d\Lambda}^{ij}\frac{\partial^2 S_{\rm int}}{\partial \phi^i \partial \phi^j}\right)~.
\ee
We have neglected the appearance of an external current in the free action, but the current drops out of the final equation if it is chosen to be orthogonal to $dP/d\Lambda$:
\be\label{eq-currentortho}
\frac{dP^{ij}}{d\Lambda}J_j = 0~.
\ee
Since $dP/d\Lambda$ only has support near the cutoff $\Lambda$, this is usually guaranteed by choosing $J$ to have support only at low momentum.

We can expand the interaction as a power series:
\be
S_{\rm int} = \sum_{m=0}^\infty \frac{1}{m!}V_{i_1\cdots i_m}(\Lambda) \phi^{i_1}\cdots\phi^{i_m}~.
\ee
Then we have the identities
\begin{align}
\frac{\partial S_{\rm int}}{\partial \phi^i} &= \sum_{m=0}^\infty \frac{1}{m!} V_{i i_1\cdots i_m}(\Lambda) \phi^{i_1}\cdots\phi^{i_m}~,\\
\frac{\partial^2 S_{\rm int}}{\partial \phi^i \partial \phi^j} &=   \sum_{m=0}^\infty \frac{1}{m!} V_{ij i_1\cdots i_m}(\Lambda) \phi^{i_1}\cdots\phi^{i_m}~.
\end{align}
The RG equation (\ref{eq-actionrg}) can be written in terms of the $V$ coefficients as
\be\label{eq-rg}
\frac{d}{d\Lambda}V_{i_1\cdots i_m}(\Lambda) = -\frac{1}{2}\left(\frac{dP}{d\Lambda}^{ij}V_{iji_1\cdots i_m}+\frac{dP}{d\Lambda}^{ij}\sum_{k=0}^m {m\choose k}V_{i i_1\cdots i_k}V_{j i_{k+1}\cdots i_m}\right)~.
\ee
There is a simple diagrammatic interpretation to this equation. The LHS represents a vertex with $m$ external legs. The first term on the RHS is a vertex with $m+2$ external legs, and two of them are contracted with the free propagator. The second term on the RHS takes a vertex with $k+1$ external lines and another with $m-k+1$ external lines and connects them by contracting one line from each vertex using a propagator.

\subsection{A Path Integral for Standard Perturbation Theory}

In cosmological perturbation theory we are given initial data $\phi_{\rm in}^i$ at the initial time $\tau_{\rm in}$ for some collection of perturbation fields. For simplicity, we will assume that these satisfy Gaussian statistics, meaning that their correlation functions can be calculated using the path integral
\be\label{eq-primordialPI}
\langle \phi^{i_1}_{\rm in} \cdots \phi^{i_n}_{\rm in}  \rangle = \int \mathcal{D}\phi_{\rm in}~ \phi^{i_1}_{\rm in} \cdots \phi^{i_n}_{\rm in} \exp\left(-\frac{1}{2} \phi_{\rm in}^i [P_{\rm in}^{-1}]_{ij} \phi_{\rm in}^j\right)~.
\ee
This is a three-dimensional Euclidean path integral over the set of all initial perturbation configurations. The matrix $P_{\rm in}$ appearing here is the initial power spectrum. Of course, in reality there will be a small amount of non-Gaussianity in the statistics at $\tau_{\rm in}$, coming from nonlinear evolution prior to $\tau_{\rm in}$ in addition to possible primordial non-Gaussianity. To account for this we can replace the exponent in (\ref{eq-primordialPI}) with a more complicated functional of the $\phi_{\rm in}$, including higher order terms.

We can use this same path integral to compute the correlation functions of the perturbations at a later time as well. Denote these late-time perturbations by $\phi^i$, suppressing reference to the time of evaluation $\tau_0$. The procedure is simply to write the late-time perturbations as functionals of the initial data, $\phi^i= \phi^i[\phi_{\rm in}]$, computed from the equations of motion, and then plug this into the path integral:
\be
\langle \phi^{i_1} \cdots \phi^{i_n} \rangle = \int \mathcal{D}\phi_{\rm in}~ \phi^{i_1}[\phi_{\rm in}] \cdots \phi^{i_n}[\phi_{\rm in}]  e^{S_0[\phi_{\rm in}]}~.
\ee
The ``free action" $S_0$ is just the same quantity which appeared in the exponent of (\ref{eq-primordialPI}).
The late-time correlation functions can be collected into a generating functional $Z[J]$:
\be
Z[J] = \int \mathcal{D}\phi_{\rm in}~ \exp\left(S_0[\phi_{\rm in}] + J_i\phi^i[\phi_{\rm in}]\right)~.
\ee
Now we note that the external current term in the exponential can be alternatively interpreted as a complicated interaction term for $\phi_{\rm in}$. By solving for $\phi^i$ in perturbation theory, we obtain a polynomial expansion for the interactions, the vertices of which are determined by integrating the equations of motion perturbatively. In SPT, we have a natural expansion for $\phi^i$:
\be
\phi^i_{\rm SPT} \equiv  K^i_{{\rm SPT}\,j}\, \phi_{\rm in}^j + \frac{1}{2} K^i_{{\rm SPT}\,jk}\,\phi_{\rm in}^j\phi_{\rm in}^k + \cdots~.
\ee
In the notation of the previous section, we would write
\be
\phi^j_{(m)} = \frac{1}{m!}  K^j_{{\rm SPT}\,i_1\cdots i_m}\, \phi_{\rm in}^{i_1}\ldots  \phi_{\rm in}^{i_m}~.
\ee
Comparison to equations (\ref{phi1oftau}) and (\ref{phi2oftau})
leads to explicit formulas for $K^i_{{\rm SPT}\,j}$ and $K^i_{{\rm SPT}\,jk}$:
\begin{align}
 K^i_{{\rm SPT}\,j} &= G^i_j(\tau_0;\tau_{\rm in})~,\\
 K^i_{{\rm SPT}\,jk} &= \int_{\tau_{\rm in}}^{\tau_0} d\tau' ~G^i_n(\tau_0;\tau') M^n_{lm}G^l_j(\tau';\tau_{\rm in})G^m_k(\tau';\tau_{\rm in})~.
\end{align}
The higher order coefficients can be computed analogously. Using these, that the $m$-point interaction vertex for $\phi_{\rm in}$ is given by
\be
V_{{\rm SPT}\, i_1\cdots i_m} \equiv  J_j K^j_{{\rm SPT}\,i_1\cdots i_{m}}~.
\ee


\subsection{RG in the Effective Theory}

We will now introduce a cutoff to the theory. First, we replace the initial power spectrum with the smoothed power spectrum in the free part of the action: $P_{\rm in}^{ij} \to W_\Lambda(i)^2 P_{\rm in}^{ij} \equiv P^{ij}$. For this section, $W_\Lambda$ is a function which is equal to $1$ for $k\lesssim \Lambda$, has a non-vanishing derivative for $k\sim \Lambda$, and then is equal to $0$ for $k\gtrsim \Lambda$. With this choice, $dP/d\Lambda$ is only nonzero over  a very small range of momenta right near the cutoff. In this way we have restricted the set of possible initial conditions for the perturbations. Second, we restrict the external current $J$ to have support only for modes $k <\Lambda$. It is important that the support of $J$ be distinct from the support of $dP/d\Lambda$. This corresponds to only allowing ourselves to ask questions about long-wavelength modes.

Having made these restrictions, we need to include additional $\Lambda$-dependent interaction terms which incorporate the effects of the UV modes that have been eliminated. We will deduce the form of these interactions by demanding that the full set is closed under the RG flow, \ref{eq-rg}. There are three classes of new terms, two of which we have already discussed. First, there are terms linear in $J$, which represent propagation. Second, there are terms which are $J$-independent and represent renormalizations of the effective initial power spectrum, as well as induced initial non-Gaussianities. The third class of terms, which we have not yet discussed, contains the terms nonlinear in $J$. Normally, terms with higher powers of an external current are not generated by RG evolution. This is because the external current is chosen to have support only over momenta for which $dP/d\Lambda = 0$, and so $J$ drops out of the RG equations as discussed above near (\ref{eq-currentortho}). However, that is not the case here. Consider the contraction of the $m$-point interaction with $dP/d\Lambda$, as it occurs in (\ref{eq-rg}):
\be
\frac{dP^{ii_1}}{d\Lambda}V_{{\rm SPT}\, i_1\cdots i_m} =\frac{dP^{ii_1}}{d\Lambda}J_j K^j_{{\rm SPT}\,i_1\cdots i_{m}}(q_1,\cdots,q_m)~.
\ee
To make things a little more clear, we will restore explicit momentum dependence to this equation and use conservation of momentum to simplify the expression. Then, by a slight abuse of notation, we have
\be
\frac{dP^{ii_1}}{d\Lambda}(q_1) V_{{\rm SPT}\, i_1\cdots i_m}(q_1,\cdots, q_m) = \int \frac{d^3q_1}{(2\pi)^3}\frac{dP^{ii_1}}{d\Lambda}(q_1)J_j\left(-\sum q_i\right) K^j_{{\rm SPT}\,i_1\cdots i_{m}}(q_1,\cdots,q_m)~.
\ee
The power spectrum derivative is only nonzero when $q_1 \approx \Lambda$, and $J$ vanishes unless $\sum q_i < \Lambda$. Except for $m=1$, these requirements can be simultaneously satisfied. Therefore $J$ does not drop out of the RG equations. Hence RG evolution is nonlinear, and we will generate terms which are nonlinear in $J$. These are generalized evolution terms for $\phi$; heuristically, they encode the effects of the stochastic source terms of the precious section.

\begin{figure*}[tb]
\begin{center}
\includegraphics[width=16 cm]{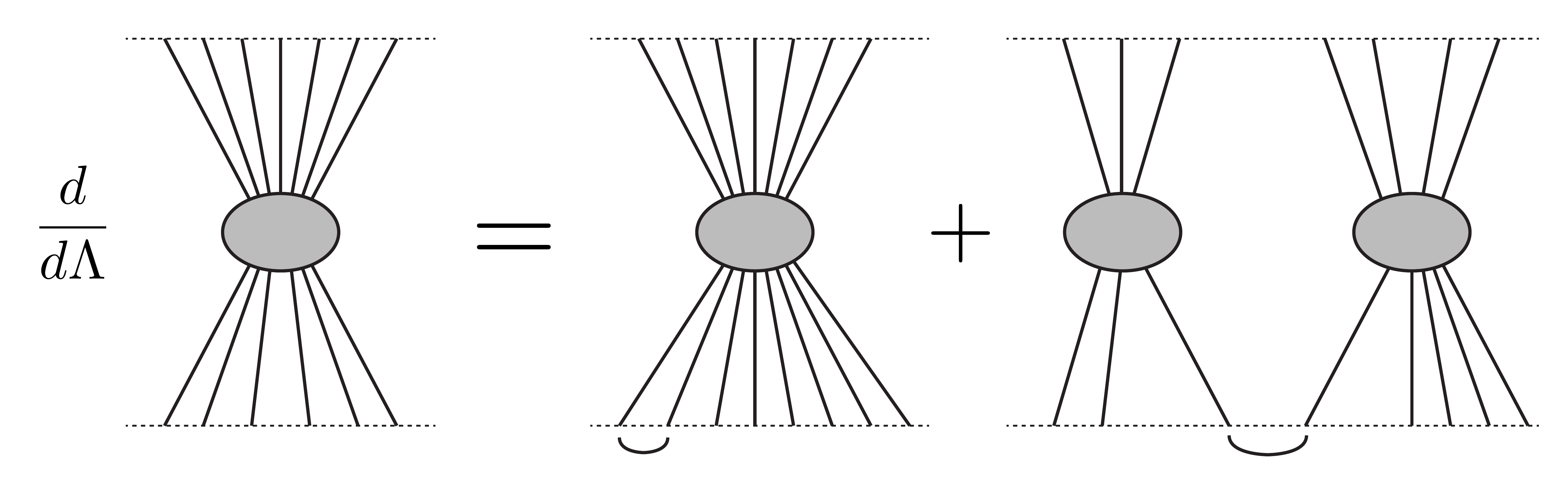}
\caption{Diagrammatic representation of the renormalization group equation (\ref{eq-rgK}). The curved bracket at the bottom represents a contraction with a factor of $dP^{ij}/d\Lambda$. The second graph on the right stands for a sum over various ways to distribute and contract the incoming lines.}
\label{RG-fig}
\end{center}
\end{figure*}

In general, we can write a series expansion for the full $m$-point interaction coefficient $V$:
\be
V_{i_1\cdots i_m} = \sum_{n=0}^\infty \frac{1}{n!}J_{j_1}\cdots J_{j_n} K^{j_1\cdots j_n}_{i_1\cdots i_m}~.
\ee
The $n=0$ and $n=1$ terms are the renormalized initial distribution function and standard time evolution, respectively, while the $n>1$ terms are the new ones. We can expand (\ref{eq-rg}) as a power series in $J$ to obtain RG equations for the $K$ coefficients. This yields
\be\label{eq-rgK}
\frac{d}{d\Lambda} K^{j_1\cdots j_n}_{i_1\cdots i_m} = -\frac{1}{2}\left(\frac{dP}{d\Lambda}^{ij}K^{j_1\cdots j_n}_{ij i_1\cdots i_m}+\frac{dP}{d\Lambda}^{ij}\sum_{k=0}^m\sum_{l=0}^n {m\choose k} {n\choose l}K^{j_1\cdots j_l}_{i i_1\cdots i_k}K^{j_{l+1}\cdots j_n}_{j i_{k+1}\cdots i_{m}}\right)~.
\ee
Since the RHS of (\ref{eq-rgK}) involves only terms with $\leq n$ raised indices, the equations can be solved order-by-order in $n$. The $n=0$ terms therefore renormalize among themselves. We immediately learn that if all $n=0$ terms vanish,  non-zero terms will never be generated by the RG. This means that a Gaussian initial distribution at all scales, and the power spectrum is unchanged.

Now we examine the $n=1$ equation:
\be
\frac{d}{d\Lambda} K^{j_1}_{i_1\cdots i_m} = -\frac{1}{2}\frac{dP}{d\Lambda}^{ij}\left[K^{j_1}_{ij i_1\cdots i_m}+\sum_{k=0}^m{m\choose k}\left( K^{j_1}_{i i_1\cdots i_k}K_{j i_{k+1}\cdots i_{m}}+K_{i i_1\cdots i_k}K^{j_1 }_{j i_{k+1}\cdots i_{m}}\right)\right]~.
\ee
If the initial perturbations are Gaussian, $K$ coefficients with zero upper indices vanish, so the first term on the RHS is the only one that survives. For the special case $m=1$, we find
\be
\frac{d}{d\Lambda} K^{j_1}_{i_1}= -\frac{1}{2} \frac{dP}{d\Lambda}^{ij}K^{j_1}_{ij i_1}~.
\ee
Since $K$ coefficients with a single upper index correspond to renormalized time evolution, we can write $K=  K_{\rm SPT} + \Delta  K$ for these, where $K_{\rm SPT}$ is defined above and is independent of $\Lambda$. Then we have
\be\label{eq-K13}
\frac{d}{d\Lambda} \Delta  K^{l}_{m}= -\frac{1}{2} \frac{dP}{d\Lambda}^{ij}( K^{l}_{{\rm SPT}\,ij m}+\Delta  K^{l}_{ijm})~.
\ee
The $\Delta  K$ factors are only nonzero because they are generated by this equation. Therefore they are suppressed relative to the $K_{\rm SPT}$ factors by $P^{ij}$. To the extent that perturbation theory is valid, this means they are small. 

The utility of this equation will be more recognizable if we compute $\langle \phi^n_{(1)} \phi^m_{(3)} \rangle_\Lambda$ in the notation of this section, where the subscript $\Lambda$ means that we take the SPT calculation and restrict the initial conditions to be long-wavelength:
\begin{align}
\langle \phi^n_{(1)} \phi^m_{(3)} \rangle_\Lambda &= \frac{1}{6} K^m_{{\rm SPT}\,jkl}\left( \langle \phi^n_{(1)} \phi_{\rm in}^j\rangle_\Lambda \langle \phi_{\rm in}^k \phi_{\rm in}^l \rangle_\Lambda+\langle \phi^n_{(1)} \phi_{\rm in}^k\rangle_\Lambda\langle \phi_{\rm in}^j \phi_{\rm in}^l \rangle_\Lambda+\langle \phi^n_{(1)} \phi_{\rm in}^l\rangle_\Lambda\langle \phi_{\rm in}^j \phi_{\rm in}^k \rangle_\Lambda\right)\\
&= \frac{1}{6} K^m_{{\rm SPT}\,jkl}\left( \langle \phi^n_{(1)} \phi_{\rm in}^j\rangle_\Lambda P^{kl}+\langle \phi^n_{(1)} \phi_{\rm in}^k\rangle_\Lambda P^{jl}+\langle \phi^n_{(1)} \phi_{\rm in}^l\rangle_\Lambda P^{jk}\right)\\
&= \frac{1}{2} K^m_{{\rm SPT}\,jkl}\langle \phi^n_{(1)} \phi_{\rm in}^l\rangle_\Lambda P^{jk}\\
&= \frac{1}{2} K^m_{{\rm SPT}\,jkl}\langle \phi^n_{(1)} \phi_{\rm in}^l\rangle P^{jk}~.
\end{align}
In the last equality we have used the fact that the momentum associated with the $n$ index is below the cutoff, so $\langle \phi^n_{(1)} \phi_{\rm in}^l\rangle_\Lambda$ is actually independent of $\Lambda$. Then all of the $\Lambda$-dependence of this expression is in the $P^{jk}$ factor. Taking the $\Lambda$-derivative and using (\ref{eq-K13}) we find
\begin{align}
\frac{d}{d\Lambda} \langle \phi^n_{(1)} \phi^m_{(3)} \rangle_\Lambda &= \frac{1}{2} K^m_{{\rm SPT}\,jkl}\langle \phi^n_{(1)} \phi_{\rm in}^l\rangle \frac{dP}{d\Lambda}^{jk}\\
&=-\langle \phi^n_{(1)} \phi_{\rm in}^l\rangle \left(\frac{d}{d\Lambda} \Delta  K^m_l + \frac{dP}{d\Lambda}^{jk}\frac{1}{2}\Delta  K^m_{jkl}\right)\\
&=-\frac{d}{d\Lambda}\left(\langle \phi^n_{(1)} \Delta  K^m_l\phi_{\rm in}^l\rangle\right) -\frac{1}{2}\Delta  K^m_{jkl}\langle \phi^n_{(1)} \phi_{\rm in}^l\rangle  \frac{dP}{d\Lambda}^{jk}
\end{align}
Since the $\Delta  K$ terms are small in perturbation theory, the second term on the RHS can be dropped as it is subdominant compared to the first term. Then this equation says that the $\Lambda$-dependence of $\langle \phi_{(1)}^n \phi_{(3)}^m \rangle$ is canceled by $\Lambda$ dependence of $\langle \phi_{(1)}^n \Delta\phi_{(3)}^m  \rangle$, where $\Delta\phi_{(3)}^m  = \Delta K^m_l\phi_{\rm in}^l$. We can think of this as a renormalization of the propagation of $\phi$.

It is tempting to identify this correction with $\Delta \phi_{{\rm L}(3)}$ from Section~\ref{sec-smoothing}, but they are not the same. Recall that $\Delta \phi_{{\rm L}(3)}$ was defined as the correction to $\phi_{{\rm L}(3)}$, which was related to $\phi_{(3)}$ be restricting every propagating state to be long-wavelength. Here we are not doing that. We are only restricting the initial conditions to be long-wavelength, but these may generate short-wavelength states spontaneously. The spontaneously generated short-wavelength states remain linear, as discussed in Section~\ref{sec-smoothing}, and so they are automatically incorporated in SPT. The correction $\Delta \phi_{(3)}$ of the present section is the correction to the SPT result (with restricted initial conditions), and so does not include these particular modes.

Now we turn to the $n=2$ equation, which represents the simplest type of generalized evolution. In that case we have
\be
\frac{d}{d\Lambda} K^{j_1 j_2}_{i_1\cdots i_m} = -\frac{1}{2}\left(\frac{dP}{d\Lambda}^{ij}K^{j_1 j_2}_{ij i_1\cdots i_m}+\frac{dP}{d\Lambda}^{ij}\sum_{k=0}^m\sum_{l=0}^2 {m\choose k} {2\choose l}K^{j_1\cdots j_l}_{i i_1\cdots i_k}K^{j_{l+1}\cdots j_2}_{j i_{k+1}\cdots i_{m}}\right)~.
\ee
The first term on the RHS is a generalized evolution term, and so will be subleading compared to terms in the sum which are not suppressed by the power spectrum. If the initial perturbations are Gaussian, the $l=0$ and $l=2$ terms in the sum on the RHS vanish. The remaining sum over $k$ vanishes for $m=0$ and $m=1$ by momentum conservation and orthogonality of $J$ and $dP/d\Lambda$. The simplest nontrivial case is $m=2$, where we find the equation reduces to 
\be
\frac{d}{d\Lambda} K^{j_1 j_2}_{i_1 i_2} = -\frac{1}{2}\left(\frac{dP}{d\Lambda}^{ij}K^{j_1 j_2}_{ij i_1 i_2}+4\frac{dP}{d\Lambda}^{ij} K^{j_1}_{i i_1}K^{ j_2}_{j  i_{2}}\right)~.
\ee
To leading order in perturbation theory, this equation is
\be
\frac{d}{d\Lambda} K^{j_1 j_2}_{i_1 i_2} = -2\frac{dP}{d\Lambda}^{ij}  K^{j_1}_{{\rm SPT}\,i i_1} K^{ j_2}_{{\rm SPT}\,j  i_{2}}~
\ee
The meaning of this equation can be illuminated by considering $\langle \phi^n_{(2)}\phi^m_{(2)}\rangle_\Lambda$:
\begin{align}
\langle \phi^n_{(2)} \phi^m_{(2)} \rangle_\Lambda &= \frac{1}{4} K^n_{{\rm SPT}\, ij} K^m_{{\rm SPT}\,kl}\left( \langle \phi^i_{\rm in} \phi_{\rm in}^j\rangle_\Lambda \langle \phi^k_{\rm in} \phi_{\rm in}^l\rangle_\Lambda +\langle \phi^i_{\rm in} \phi_{\rm in}^k\rangle_\Lambda \langle \phi^j_{\rm in} \phi_{\rm in}^l\rangle_\Lambda+\langle \phi^i_{\rm in} \phi_{\rm in}^l\rangle_\Lambda \langle \phi^k_{\rm in} \phi_{\rm in}^j\rangle_\Lambda\right)\\
&=\frac{1}{4}K^n_{{\rm SPT}\, ij} K^m_{{\rm SPT}\,kl}\left(P^{ij}P^{kl} + P^{ik} P^{jl} + P^{il}P^{kj}\right)\\
&=\frac{1}{2}K^n_{{\rm SPT}\, ij} K^m_{{\rm SPT}\,kl}P^{ik} P^{jl}+\langle \phi_{(2)}^n\rangle_\Lambda\langle \phi_{(2)}^m\rangle_\Lambda~.
\end{align}
The second term vanishes in the theory of LSS based on the explicit form of $K_{\rm SPT}$, which we can see in Appendix~\ref{sec-pertcalc}. To leading order in perturbation theory, the $\Lambda$-derivative of the first term is
\begin{align}
\frac{d}{d\Lambda}\langle \phi^n_{(2)} \phi^m_{(2)} \rangle_\Lambda  &=K^n_{{\rm SPT}\, ij} K^m_{{\rm SPT}\,kl}P^{ik} \frac{dP}{d\Lambda}^{jl}\\
&= -\frac{1}{2} P^{ik} \frac{d}{d\Lambda} K^{nm}_{ik}\\
&= -\frac{1}{2} \frac{d}{d\Lambda}\left(P^{ik} K^{nm}_{ik}\right)+\frac{1}{2} \frac{dP}{d\Lambda}^{ik} K^{nm}_{ik} \\
&= -\frac{1}{2} \frac{d}{d\Lambda}\left(P^{ik} K^{nm}_{ik} + 2K^{nm}\right) ~.
\end{align}
In the last line we have used another of the RG equations. Notice that the $\langle \phi_{(2)} \phi_{(2)}\rangle_\Lambda$ contribution to the correlation function has its $\Lambda$-dependence canceled by new types of terms not present in the classical theory. The second term, in particular, is strange because it has no lowered indices. That means it comes from a field-independent (but $J$-dependent) interaction in the effective action. 

Let us take a moment to examine the RG equation associated with such terms:
\be
\frac{d}{d\Lambda} K^{j_1\cdots j_n} = -\frac{1}{2}\left(\frac{dP}{d\Lambda}^{ij}K^{j_1\cdots j_n}_{ij}+\frac{dP}{d\Lambda}^{ij}\sum_{l=0}^n  {n\choose l}K^{j_1\cdots j_l}_{i }K^{j_{l+1}\cdots j_n}_{j }\right)~.
\ee
We will restrict our attention to $n=1$ and $n=2$, since those are the terms which are relevant for the one- and two-point functions, and for simplicity we continue to assume a Gaussian initial power spectrum.
For $n=1$, the equation becomes
\be
\frac{d}{d\Lambda} K^{j_1} = -\frac{1}{2}\frac{dP}{d\Lambda}^{ij}K^{j_1}_{ij}~.
\ee
We can once more write $K^{j_1}_{ij} =  K^{j_1}_{{\rm SPT}\,ij} + \Delta  K^{j_1}_{ij}$. In the theory of LSS, the term with $  K_{\rm SPT}$ vanishes due to momentum conservation and the explicit form of $K_{\rm SPT}$. Therefore the only contribution is from $ \Delta  K^{j_1}_{ij}$. So $K^{j_1}$ is suppressed by two powers of $P^{ij}$.

Now we turn to $n=2$:
\be
\frac{d}{d\Lambda} K^{j_1 j_2} = -\frac{1}{2}\left(\frac{dP}{d\Lambda}^{ij}K^{j_1 j_2}_{ij}+2\frac{dP}{d\Lambda}^{ij} K^{j_1}_{i }K^{ j_2}_{j }\right)~.
\ee
The second term actually vanishes by momentum conservation and the orthogonality of $dP/d\Lambda$ and $J$. Then $K^{j_1 j_2}$ is also generated at second order in $P$.


\subsection{Renormalization of the One-Point Function and Power Spectrum}
We can use the above formalism to compute the one-point function and two-point function by taking derivatives of $Z$. The first derivative of $Z$ is
\be\label{eq-zderiv}
\frac{\partial Z}{\partial J_i} = Z[J] \sum_{m=0}^\infty  \sum_{n=0}^\infty \frac{1}{m!n!}J_{j_1}\cdots J_{j_n} K^{i j_1\cdots j_n}_{i_1\cdots i_m}\langle\phi_{\rm in}^{i_1}\cdots\phi_{\rm in}^{i_m}\rangle^J_\Lambda~,
\ee
where $\langle\cdots \rangle^J_\Lambda$ denotes the expectation value in the presence of nonzero $J$ (as well as the cutoff $\Lambda$). This means that the expectation value of $\phi$ is
\be\label{eq-onepoint}
\langle \phi^i \rangle_\Lambda = \frac{1}{Z[0]} \left. \frac{\partial Z}{\partial J_i}\right|_{J=0} = \sum_{m=0}^\infty  \frac{1}{m!}K^{i}_{i_1\cdots i_m}\langle\phi_{\rm in}^{i_1}\cdots\phi_{\rm in}^{i_m}\rangle_\Lambda~.
\ee
None of the nonlinear terms in $J$ are involved in this computation. Generally speaking, the $n$-point function requires knowledge of the $J^k$ terms in the action for $k\leq n$. So the one-point function is based purely on the renormalized evolution defined by the interaction terms linear in $J$. Also note that all of these expectation values are taken with $J=0$, so that they are Gaussian expectation values if the initial distribution is Gaussian, using the cut-off initial power spectrum $W_\Lambda(i)^2P_{\rm in}^{ij}$. In this special case, only the even terms in (\ref{eq-onepoint}) are nonvanishing.

To compute the two-point function, we take another derivative of (\ref{eq-zderiv}), divide by $Z$, and set $J=0$:
\be\begin{split}
\langle \phi^i \phi^j \rangle_\Lambda &= \frac{1}{Z[0]} \left. \frac{\partial^2 Z}{\partial J_i \partial J_j}\right|_{J=0} \\
&= \sum_{m=0}^\infty  \sum_{k=0}^m \frac{1}{m!} {m\choose k} K^{i }_{i_1\cdots i_k}K^{j }_{i_{k+1}\cdots i_m}\langle\phi_{\rm in}^{i_1}\cdots\phi_{\rm in}^{i_m}\rangle_\Lambda \\
&\qquad + \sum_{m=0}^\infty \frac{1}{m!}  K^{i j}_{i_1\cdots i_m}\langle\phi_{\rm in}^{i_1}\cdots\phi_{\rm in}^{i_m}\rangle_\Lambda+\langle\phi^i\rangle_\Lambda\langle \phi^j\rangle_\Lambda~.
\end{split}
\ee
The third term is the disconnected piece, and the first term is the answer one would expect if the effective theory simply consisted of renormalized classical propagation of $\phi$, while the second term is the generalized evolution.

For simplicity, let us set the disconnected piece to zero and assume Gaussian initial perturbations. Then up to second order in $P^{ij}$ we have
\be\begin{split}
\langle \phi^i \phi^j \rangle 
&= \langle\phi_{(1)}^i\phi_{(1)}^j\rangle + \left[ \langle\phi_{(1)}^i\phi_{(3)}^j\rangle_\Lambda+ \langle\phi_{(1)}^i \Delta  K^j_{k}\,\phi_{\rm in}^k\rangle_\Lambda + (i \leftrightarrow j)\right] \\
&\qquad\qquad +\left[ \langle\phi_{(2)}^i\phi_{(2)}^j\rangle_\Lambda +  K^{ij}+\frac{1}{2}  K^{i j}_{i_1i_2}P^{i_1 i_2}\right]~.
\end{split}
\ee
The first term is the linear evolution, and is $\Lambda$-independent. The bracketed collections of terms are each $\Lambda$-independent at this order in perturbation theory based on the analysis above. This expression can be compared with (\ref{eq-SPTpowerspectrum}) in standard perturbation theory, or (\ref{eq-phiLpowerspectrum}) in the smoothing formalism.

Our analysis of the path integral has been fairly formal, but nevertheless extremely instructive. 
This approach to cosmological perturbation theory reveals which collections of terms must have vanishing $\Lambda$-dependence, as well as indicating what kinds of structures are generated by renormalization. It is also able to automatically incorporate the effects of short-wavelength modes that are generated by interactions of the long-wavelength modes (and are therefore in the linear regime). The appearance of integration kernels $K$ with multiple upper indices is novel, representing effects directly attributable to these generated short-wavelength modes. It would be interesting to connect these with the statistics of a stochastic source term in the equations of motion.


\section{Conclusions and Further Directions}\label{sec-conclusion}

The application of effective field theory ideas to the evolution of large-scale structure involves a number of subtle issues. Unlike in quantum field theory, here the underlying model is a classical field theory with probabilistic initial conditions, in which modes at all wavelengths can propagate over large distances. Fortunately, in the real world it is short-wavelength modes that are nonlinear, allowing us to construct a perturbation theory for the long-wavelength modes.

In this paper we analyzed two methods of deriving such a theory. The first approach, which proceeds by smoothing the fields, most closely resembles previous work in the subject. We were able to carefully derive consistent expressions for the long-wavelength modes by first expressing the short-wavelength parts as a Taylor expansion in the long-wavelength background. This makes the dependence on nonlinear effects very explicit. There is a technical speed bump inherent in this technique, however, due to the ability of interactions between long-wavelength modes to create perturbative short-wavelength modes, which the smoothing procedure automatically squelches.

Our other method starts with a path integral over initial perturbations, and uses the renormalization group to derive conditions obeyed by correlation functions. This approach is quite general and flexible, and naturally accounts for the effects of perturbative short-wavelength modes. Further investigation will be required to see how practical this technique is for calculations beyond what is shown in Appendix~\ref{sec-pertcalc} (although we see no reason why it shouldn't be).

The main open question is that of predictivity. In both methods a huge number of effective interactions are generated. Much of the power of EFT as used in quantum field theory comes from the reduction of the possible number of terms due to symmetries. It has been argued in~\cite{Carrasco:2012cv,Hertzberg:2012qn} that only a certain collection of terms are present in the effective theory. However, it is not yet clear in the formalisms we discuss precisely how the symmetries act to simplify the equations. In particular, the nonlocal-in-time interactions of the smoothed picture are qualitatively different from the types of interactions we are used to considering, and further study may illuminate some unexpected properties.


\acknowledgments
We thank Clifford Cheung and Mark Wise for helpful conversations. This research is funded in part by DOE grant No.~DE-FG02-92ER40701, and by the Gordon and Betty Moore Foundation through Grant No.~776 to the Caltech Moore Center for Theoretical Cosmology and Physics. The work of SL is supported by a John A. McCone Postdoctoral Fellowship.


\appendix


\section{Explicit Calculations}\label{sec-pertcalc}

In this appendix we perform some explicit calculations in both SPT and the EFT of LSS in a simple setting. We will use  a single-component, nonrelativistic, rotation-free dark matter fluid moving in an Einstein-de Sitter background, subject to Newtonian gravitational attraction. Extensions to multi-component fluids and the full theory of general relativity are straightforward, but distracting. All of the fundamental conceptual issues are present already in this simple calculation.


\subsection{Standard Perturbation Theory}

In an Einstein-de Sitter cosmology---that is, a flat, matter-dominated FRW universe---the scale factor is given by $a(\tau) = (\tau/\tau_0)^2$, the conformal Hubble factor is $\mathcal{H} = a^{-1}da/d\tau = 2/\tau$, and the equations of motion for a dark matter fluid simplify to 
\begin{align}
0&=\partial_\tau \delta(\tau,{\bf {\bf k}}) + \theta(\tau,{\bf {\bf k}}) + \int \frac{d^3q}{(2\pi)^3} \frac{{\bf {\bf k}} \cdot {\bf q}}{q^2} \delta(\tau,{\bf {\bf k}}-{\bf q})\theta(\tau, {\bf q}) ~,  \\
0&=\partial_\tau \theta(\tau,{\bf {\bf k}}) + \mathcal{H} \theta(\tau,{\bf {\bf k}}) + \frac{3}{2}\mathcal{H}^2 \delta(\tau,{\bf {\bf k}})
+ \int \frac{d^3q}{(2\pi)^3} \frac{k^2 {\bf q}\cdot({\bf {\bf k}}-{\bf q})}{2q^2 ({\bf {\bf k}}-{\bf q})^2} \theta(\tau,{\bf {\bf k}}-{\bf q})\theta(\tau,{\bf q}).
\end{align}

At first order, we drop the nonlinear terms and solve the linearized equations using a Green function:
\be\label{eq-FirstOrderMatrix}
\begin{pmatrix}
\delta_{(1)}(\tau,{\bf {\bf k}}) \\ \theta_{(1)}(\tau,{\bf {\bf k}}) 
\end{pmatrix}
 = G(\tau;\tau_{\rm in}) \begin{pmatrix}
\delta_{\rm in}({\bf {\bf k}}) \\ \theta_{\rm in}({\bf {\bf k}}) 
\end{pmatrix}~.
\ee
Here $\delta_{\rm in}$ and $\theta_{\rm in}$ are initial conditions at the initial time $\tau_{\rm in}$, and $G(\tau,\tau_{\rm in})$ is a retarded Green function for the linearized system:
\be\label{eq-EdSGreens}
G(\tau_1;\tau_2) = \begin{pmatrix} \frac{3\tau_1^5 + 2\tau_2^5}{5\tau_1^3 \tau_2^2} & \frac{-\tau_1^5 + \tau_2^5}{5\tau_1^3 \tau_2} \\ -\frac{6(\tau_1^5 - \tau_2^5)}{5\tau_1^4 \tau_2^2} & \frac{2 \tau_1^5 + 3\tau_2^5}{5\tau_1^4 \tau_2}  \end{pmatrix} \Theta(\tau_1 - \tau_2).
\ee
Note that for $\tau \gg \tau_{\rm in}$ we have $\delta_{(1)} \propto \tau^2 \propto a$, which is the dominant growth function behavior familiar from standard perturbation theory. By keeping careful track of both $\theta$ and $\delta$ we are effectively including the subdominant mode as well.

The method of SPT is to incorporate nonlinearities perturbatively by substituting the linear solution into the nonlinear terms and treating them as a source for the second-order terms. Then we have
\be\label{eq-SecondOrderMatrix}
\begin{pmatrix}
\delta_{(2)}(\tau,{\bf k}) \\ \theta_{(2)}(\tau,{\bf k}) 
\end{pmatrix}
= - \int_{\tau_{\rm in}}^\tau d\tau' \int \frac{d^3q}{(2\pi)^3}~G(\tau;\tau')\begin{pmatrix}
 \frac{{\bf k} \cdot {\bf q}}{q^2}\delta_{(1)}(\tau',{\bf k}-{\bf q})\theta_{(1)}(\tau', {\bf q})  \\   \frac{k^2 {\bf q}\cdot({\bf k}-{\bf q})}{2q^2 ({\bf k}-{\bf q})^2} \theta_{(1)}(\tau',{\bf k}-{\bf q})\theta_{(1)}(\tau',{\bf q})
\end{pmatrix}.
\ee
This procedure can be repeated to compute higher-order perturbations. For instance, at the next order we have 
\begin{align}
\begin{pmatrix}
\delta_{(3)}(\tau,{\bf k}) \\ \theta_{(3)}(\tau,{\bf k}) 
\end{pmatrix} 
 =& - \int_{\tau_{\rm in}}^\tau d\tau' \int \frac{d^3q}{(2\pi)^3}~G(\tau;\tau') \nonumber\\
& \qquad\qquad \times \begin{pmatrix}\frac{{\bf k} \cdot {\bf q}}{q^2}\left[\delta_{(1)}(\tau',{\bf k}-{\bf q})\theta_{(2)}(\tau', {\bf q}) +\delta_{(2)}(\tau',{\bf k}-{\bf q})\theta_{(1)}(\tau', {\bf q}) \right] \\   \frac{k^2 {\bf q}\cdot({\bf k}-{\bf q})}{q^2 ({\bf k}-{\bf q})^2} \theta_{(1)}(\tau',{\bf k}-{\bf q})\theta_{(2)}(\tau',{\bf q})
\end{pmatrix}.
\label{eq-ThirdOrderMatrix}
\end{align}
The expansion is tedious but straightforward to compute.

Before continuing, it will be useful to make contact with the notation of the body of the paper.  The perturbations $\delta$ and $\theta$ are regarded as components of the perturbation field $\phi^i$, where $i=\delta$ or $i=\theta$ (for added clarity we label the momentum separately, and not as part of the Latin index). Then (\ref{eq-FirstOrderMatrix}) can be replaced by
\be\label{eq-phi1}
\phi^i_{(1)}(\tau, {\bf k}) = G^i_j(\tau;\tau_{\rm in}) \phi_{\rm in}^j({\bf k})~.
\ee
The nonlinear terms in the equations of motion can be incorporated into the pair of matrices
\begin{align}
M^\delta_{ij}({\bf k}_1; {\bf k}_2, {\bf k}_3) &= \begin{pmatrix} 0 & - \frac{{\bf k}_1 \cdot {\bf k}_3}{k_3^2} \\  - \frac{{\bf k}_1 \cdot {\bf k}_2}{k_2^2}& 0  \end{pmatrix}~,\\
M^\theta_{ij}({\bf k}_1;{\bf k}_2,{\bf k}_3) &= \begin{pmatrix} 0 &0 \\0 & -\frac{k_1^2 ({\bf k}_2\cdot {\bf k}_3)}{k_2^2 k_3^2} \end{pmatrix}~,
\end{align}
so that (\ref{eq-SecondOrderMatrix}) and (\ref{eq-ThirdOrderMatrix}) become, respectively,
\be\label{eq-phi2}
\phi^i_{(2)}(\tau, {\bf k}) = \frac{1}{2}\int_{\tau_{\rm in}}^\tau d\tau' \int \frac{d^3q}{(2\pi)^3}~G^i_j(\tau;\tau') M^j_{lm}({\bf k};{\bf k}-{\bf q},{\bf q})\phi^l_{(1)}(\tau',{\bf k}-{\bf q})\phi^m_{(1)}(\tau',{\bf q})~,
\ee
and 
\be\label{eq-phi3}
\phi^i_{(3)}(\tau, {\bf k}) = \int_{\tau_{\rm in}}^\tau d\tau' \int \frac{d^3q}{(2\pi)^3}~G^i_j(\tau;\tau') M^j_{lm}({\bf k};{\bf k}-{\bf q},{\bf q})\phi^l_{(1)}(\tau',{\bf k}-{\bf q})\phi^m_{(2)}(\tau',{\bf q})~.
\ee

The correlation function we will use as our example is
\be
\langle \phi^i_{(1)}(\tau_1, {\bf k}_1) \phi^j_{(3)}(\tau_2, {\bf k}_2)\rangle~,
\ee
which occurs in the SPT expansion of 
\be
\langle \phi^i(\tau_1, {\bf k}_1) \phi^j(\tau_2,{\bf k}_2)\rangle~.
\ee
We just have to multiply $\phi_{(1)}$ and $\phi_{(3)}$ above and take the expectation value using Wick's theorem (assuming Gaussian initial conditions). The result will depend on the linear power spectrum
\be
\langle \phi^i_{(1)}(\tau_1,{\bf k}_1) \phi^j_{(1)}(\tau_2,{\bf k}_2) \rangle \equiv P_{(11)}^{ij}(\tau_1,\tau_2 | {\bf k}_1)(2\pi)^3 \delta^{(3)}({\bf k}_1 + {\bf k}_2)~.
\ee
The presence of the Dirac $\delta$-function is a result of translation invariance, and in addition rotational invariance implies that $P_{(11)}^{ij}({\bf k})$ is a function only of $k^2$. These results produce numerous simplifications, in particular that $\langle \phi^i_{(2)}\rangle = 0$. Doing a little algebra, we find that 
\be\label{eq-P13}
\begin{split}
\langle \phi^i_{(1)}(\tau_1,{\bf k}_1) &\phi^j_{(3)}(\tau_2, {\bf k}_2) \rangle=(2\pi)^3 \delta^{(3)}({\bf k}_1 + {\bf k}_2)
\int_{\tau_{\rm in}}^{\tau_2} d\tau' \int_{\tau_{\rm in}}^{\tau'}d\tau'' \int\frac{d^3q}{(2\pi)^3} G^j_k(\tau_2; \tau')\\
&M^k_{lm}({\bf k}_2;{\bf q},{\bf k}_2-{\bf q})G^m_n(\tau';\tau'') M^n_{op}({\bf k}_2-{\bf q};-{\bf q},{\bf k}_2)
P_{(11)}^{ip}(\tau_1,\tau''|k_1)P_{(11)}^{lo}(\tau',\tau''|q) ~.
\end{split}
\ee
The overall factor of $(2\pi)^3 \delta^{(3)}({\bf k}_1 + {\bf k}_2)$ is present because translation invariance is respected at each order in perturbation theory. Here $q$ is the loop momentum, which still must be integrated over.


\subsection{Accounting for the Cutoff}

Now we will discuss the two approaches to regulating $\langle\phi_{(1)} \phi_{(3)} \rangle$, before discussing how the $\Lambda$-dependence of the result is canceled by new interactions. First, the smoothing method of Section~\ref{sec-smoothing} demands that we replace $\phi_{(1)}$ and $\phi_{(3)}$ with $\phi_{{\rm L}(1)}$ and $\phi_{{\rm L}(3)}$. The result is
\be
\begin{split}
\langle \phi^i_{{\rm L}(1)}&(\tau_1,{\bf k}_1) \phi^j_{{\rm L}(3)}(\tau_2, {\bf k}_2) \rangle=(2\pi)^3 \delta^{(3)}({\bf k}_1 + {\bf k}_2)\\
&\times\int_{\tau_{\rm in}}^{\tau_2} d\tau' \int_{\tau_{\rm in}}^{\tau'}d\tau'' \int\frac{d^3q}{(2\pi)^3}~ \bigg[G^j_k(\tau_2; \tau')W_\Lambda(k_2)M^k_{lm}({\bf k}_2;{\bf q},{\bf k}_2-{\bf q})G^m_n(\tau';\tau'')\\
& ~\times W_\Lambda(|{\bf k}_2 - {\bf q}|)M^n_{op}({\bf k}_2-{\bf q};-{\bf q},{\bf k}_2)W_\Lambda(k_1)^2 P_{(11)}^{ip}(\tau_1,\tau''|k_1)W_\Lambda (q)^2P_{(11)}^{lo}(\tau',\tau''|q) \bigg]~.
\end{split}
\ee
Notice that there are two nontrivial constraints on the loop integral. The two constraints $|{\bf k}_2 - {\bf q}|<\Lambda$ and $q < \Lambda$ are awkward to satisfy simultaneously. However, if we ignore the constraint on $|{\bf k}_2 - {\bf q}|$ then we only make an error of order $k_2 / \Lambda$, which is small for us.

The alternative approach is the RG path integral of Section~\ref{sec-pathintegral}. There we are instructed to only place a cutoff on the initial conditions:
\be
\begin{split}
\langle \phi^i_{(1)}(\tau_1,{\bf k}_1) &\phi^j_{(3)}(\tau_2, {\bf k}_2) \rangle_\Lambda =(2\pi)^3 \delta^{(3)}({\bf k}_1 + {\bf k}_2)\times\\
&\int_{\tau_{\rm in}}^{\tau_2} d\tau' \int_{\tau_{\rm in}}^{\tau'}d\tau'' \int\frac{d^3q}{(2\pi)^3}~ \bigg[G^j_k(\tau_2; \tau')M^k_{lm}({\bf k}_2;{\bf q},{\bf k}_2-{\bf q})G^m_n(\tau';\tau'')\times\\
& ~~~M^n_{op}({\bf k}_2-{\bf q};-{\bf q},{\bf k}_2)W_\Lambda(k_1)^2 P_{(11)}^{ip}(\tau_1,\tau''|k_1)W_\Lambda (q)^2P_{(11)}^{lo}(\tau',\tau''|q) \bigg]~.
\end{split}
\ee
Unlike the smoothing prescription, here do not place a bound on $|{\bf k}_2 - {\bf q}|$. The other difference is a missing factor of $W_\Lambda(k_2)$, but this factor is redundant since ${\bf k}_2$ is an external momentum which must be small.  This integral is more straightforward to compute, but, as mentioned above, until we are ready to compute to the level of $k_2/\Lambda$ precision there is no real difference. Either computation is well-approximated by (\ref{eq-P13}) with a hard momentum cutoff at $q=\Lambda$.

We continue the calculation be performing the angular part of the integral over ${\bf q}$ in (\ref{eq-P13}). Defining the matrix $\Sigma$ by
\be
\Sigma^k_p(\tau';\tau''|{\bf k})  = \int \frac{d^3q}{(2\pi)^3}~M^k_{lm}({\bf k};{\bf q},{\bf k}-{\bf q})G^m_n(\tau';\tau'') M^n_{op}({\bf k}-{\bf q};-{\bf q},{\bf k})P_{(11)}^{lo}(\tau',\tau''|q)~,
\ee
we have
\be
\Sigma^k_p(\tau';\tau''|{\bf k})  = \frac{k^3}{4\pi}\int dr~\left[-\frac{2}{3}P_{(11)}^{\theta\theta}(\tau',\tau''|kr)G^k_p(\tau';\tau'') + \frac{1}{2}G^\theta_\delta(\tau';\tau'')\Pi^k_p \right] ~,
\ee
where $q \equiv kr$. The entries of the $\Pi$ matrix are 
\begin{align}\label{eq-pi}
\Pi^\delta_\delta &= \left(1+r^2+\frac{1}{4r}\left(1-r^2\right)^2 \log \frac{(1-r)^2}{(1+r)^2}\right)P_{(11)}^{\delta\theta}(\tau',\tau''|kr)~,\\
\label{eq-pi2}
\Pi^\delta_\theta &= -\left(-\frac{5}{3}r^2+r^4+\frac{r}{4}\left(1-r^2\right)^2 \log \frac{(1-r)^2}{(1+r)^2}\right)P_{(11)}^{\delta\delta}(\tau',\tau''|kr)~,\\
\label{eq-pi3}
\Pi^\theta_\delta &= \left(r^{-2}-\frac{5}{3}+\frac{1}{4r^3}\left(1-r^2\right)^2 \log \frac{(1-r)^2}{(1+r)^2}\right)P_{(11)}^{\theta\theta}(\tau',\tau''|kr)~,\\
\label{eq-pi4}
\Pi^\theta_\theta &= - \left(1+r^2+\frac{1}{4r}\left(1-r^2\right)^2 \log \frac{(1-r)^2}{(1+r)^2}\right)P_{(11)}^{\theta\delta}(\tau',\tau''|kr)~.
\end{align}
Hence (\ref{eq-P13}) can be rewritten succinctly as 
\be
\langle \phi^i_{(1)}(\tau_1, {\bf k}) \phi^j_{(3)}(\tau_2, -{\bf k}) \rangle' =\int_{\tau_{\rm in}}^\tau d\tau' \int_{\tau_{\rm in}}^{\tau'}d\tau''~G^j_k(\tau_2; \tau')\Sigma^k_p(\tau';\tau''|{\bf k})P_{(11)}^{ip}(\tau_1,\tau''| k) ~.
\ee
Here the $'$ means that we have dropped the overall $\delta$-function. In an Einstein-de Sitter universe, we can use (\ref{eq-EdSGreens}) to do the time integrations analytically. The result is simple in the limit $\tau_1=\tau_2\equiv \tau \gg \tau_{\rm in}$ if we also isolate the large $r$ part of (\ref{eq-pi}-\ref{eq-pi4}). We find that
\be\label{eq-P13simple}
\langle \phi^i_{(1)}(\tau, {\bf k}) \phi^j_{(3)}(\tau, -{\bf k}) \rangle' \approx  \frac{1}{5}\frac{k^2}{4\pi^2}\begin{pmatrix}
-\frac{61}{63}& \frac{61}{63}\mathcal{H}\\
 3\mathcal{H} &  -3\mathcal{H}^2
\end{pmatrix}P_{(11)}^{\delta\delta}(\tau,\tau | {\bf k}) \int_{q\gg k} dq\,P_{(11)}^{\delta\delta}(\tau,\tau | q)~.
\ee
It is at this point that we introduce a cutoff $\Lambda$ in the integral over $q$. Approximating this integral by the contribution for $q$ near $\Lambda$ is valid up to corrections of order $k/\Lambda$.

The $\Lambda$-dependence of this correlation function needs to be canceled against the $\Lambda$-dependence of either $ \langle\phi_{{\rm L}(1)}^i \Delta \phi_{{\rm L}(3)}^j\rangle$ or $ \langle\phi_{(1)}^i \Delta  K^j_{k}\,\phi_{\rm in}^k\rangle_\Lambda$, depending on which approach we use. In this case, to the level of approximation to which we are working, they are the same. We will use the smoothing picture here because it gives us more information. The formal expression for $\Delta \phi^i_{L(3)}$ is (\ref{eq-deltaphi3}), which has several terms featuring different combinations of short and long wavelength fields. 

When we look at the expectation value $\langle\phi_{{\rm L}(1)}^i \Delta \phi_{{\rm L}(3)}^j\rangle$ there are some simplifications due to momentum conservation and Gaussian statistics. Because of Gaussian statistics, the expectation value reduces to a sum of products of pairs of expectation values. Because of momentum conservation, the long-wavelength fields can only be paired with the long-wavelength fields. Therefore terms with an odd number of $\phi_{{\rm L}(1)}$ factors when fully expanded will vanish. After some algebra, this leaves us with
\be
\langle\phi_{{\rm L}(1)}^i \Delta \phi_{{\rm L}(3)}^j\rangle = \langle \phi_{(1)}^i \Delta K^j_{k}\phi_{\rm in}^k\rangle_\Lambda\,,
\ee
where
\be
\Delta K^j_{k} = \int_{\tau_{\rm in}}^\tau d\tau'\int^{\tau'}_{\tau_{\rm in}}d\tau'' ~G^j_p(\tau;\tau') M^p_{lm} \phi_{{\rm S}0}^l \left[G^m_n(\tau';\tau'')M^n_{op} \phi_{{\rm S}0}^p + \frac{\partial \phi_{\rm S}^m(\tau')}{\partial \phi_{\rm L}^o(\tau'')}\right]G^o_k(\tau'',\tau_{\rm in})~.
\ee
We see that the form of the correction is identical to what was expected based on the analysis from Section~\ref{sec-pathintegral}. Alternatively, we can write this correction as 
\be
\langle\phi_{{\rm L}(1)}^i \Delta \phi_{{\rm L}(3)}^j\rangle' = \int^\tau_{\tau_{\rm in}} d\tau'~G^j_k(\tau; \tau')C^k_l(\tau'|k)P_{(11)}^{il}(\tau,\tau' | k)~,
\ee
with
\be
C^k_l(\tau'|k) =\int^{\tau'}_{\tau_{\rm in}}d\tau'' ~M^k_{qm} \phi_{{\rm S}0}^q \left(G^m_n(\tau';\tau'')M^n_{op} \phi_{{\rm S}0}^p + \frac{\partial \phi_{\rm S}^m(\tau')}{\partial \phi_{\rm L}^o(\tau'')}\right)\left[G(\tau';\tau'')^{-1}\right]_l^o~.
\ee
In perturbation theory, this is same result as one would have gotten from a term
\be
C^k_l ( \tau' | k) \phi^l({\bf k})
\ee
added to the equation of motion for $\phi$. 

In the theory of LSS, it has been argued~\cite{Baumann:2010tm,Carrasco:2012cv} that the terms one should add to the equations of motion represent the parameters of an imperfect fluid. In particular, to lowest order in $k/\Lambda$, the coefficients of the $C$ matrix determine the sound speed, viscosity, and heat conduction coefficients in the fluid:
\be
C(\tau | k) =\begin{pmatrix}
\chi^\delta & \chi^\theta \\
 k^2 c_s^2 &  -k^2\frac{c_v^2}{\mathcal{H}}
\end{pmatrix}~.
\ee
These four coefficients are expected on the basis of thermodynamic considerations. We should note that the $\Lambda$-dependence of (\ref{eq-P13simple}) cannot be fully accounted for with the sound speed and viscosity coefficients, $c_s^2$ and $c_v^2$ alone. Nonzero heat conduction terms are also required. This agrees with the recent result of~\cite{Mercolli:2013bsa}.


\bibliographystyle{utcaps}
\bibliography{EFTRefs}
\end{document}